\documentclass[%
 reprint,
 amsmath,amssymb,
 aps,
]{revtex4-1}

\usepackage{graphicx}% Include figure files
\usepackage{dcolumn}% Align table columns on decimal point
\usepackage{bm}% bold math
\usepackage{color}
\begin{document}

\title{Healing of polymer interfaces: Interfacial dynamics, entanglements, and strength}
\author{Ting Ge$^1$, Mark O. Robbins$^1$, Dvora Perahia$^2$ and Gary S. Grest$^3$}
\affiliation{$^1$Department of Physics and Astronomy, Johns Hopkins University, Baltimore, MD 21218 USA}
\affiliation{$^2$Department of Chemistry, Clemson University, Clemson, SC 29634 USA}
\affiliation{$^3$Sandia National Laboratories, Albuquerque, NM 87185 USA}

\date{\today}
\begin{abstract}

Self-healing of polymer films often takes place as the molecules diffuse across a damaged region, above their melting temperature.
Using molecular dynamics simulations we probe the healing of polymer films and compare the results with those obtained for thermal welding of homopolymer slabs.
These two processes differ from each other in their interfacial structure since damage leads to increased polydispersity and more short chains.
 A polymer sample was cut into two separate films that were then held
together in the melt state. The recovery of the damaged film was followed as time elapsed and polymer molecules diffused across the interface.
The mass uptake and formation of entanglements, as obtained from primitive path analysis, are extracted and correlated with the interfacial strength obtained from shear simulations. We find that the diffusion across the interface is significantly faster in the damaged film compared to welding because
of the presence of short chains.
Though interfacial entanglements increase more rapidly for the damaged films, a large fraction of these entanglements are near chain ends. As a result, the interfacial strength of the healing film increases more slowly than for welding.
For both healing and welding, the interfacial strength saturates as the bulk entanglement density is recovered across the interface.
However, the saturation strength of the damaged film is below the bulk strength for the polymer sample.
At saturation, cut chains remain near the healing interface.
They are less entangled and as a result they mechanically weaken the interface. When the strength of the interface saturates, the number of interfacial entanglements scales with the corresponding bulk entanglement density. Chain stiffness increases the density of entanglements, which increases the strength of the interface.
Our results show that a few entanglements across the interface are sufficient to resist interfacial chain pullout and enhance the mechanical strength.

\begin{description}
\item[PACS numbers: 82.35.Gh,81.20.Vj,68.35.Fx,83.10.Mj]
\end{description}
\end{abstract}
\maketitle

\section{\label{sec:intro}Introduction}
Polymer interfaces are in the core of numerous devices.
Understanding the structure and mechanical response of polymers in the interfacial region is a key to their utilization as coatings and adhesives, as well as in nano-fabrication and many other applications \cite{wool95,haward97,jone99}.
One important mechanism for strengthening interfaces is interdiffusion of polymers between opposing sides\cite{jud81,wool81, prager81,prager83,kim83,adolf85,mikos89,russell93,wool95,wool08,blaiszik10}.
This diffusion is an integral part of making welded joints.
It also underlies the remarkable ability of polymers to heal after fracturing
or being scratched.
Both welding and healing processes often involve heating the polymer interface into the melt state for a certain time period and then quenching below the glass transition temperature $T_g$ where the translational motion of the molecules is essentially arrested.
The goal is to achieve an interface with strength equal to that of the bulk.
The main difference between the two processes is the composition of the interface since homogeneity of welded interfaces can be well controlled whereas damaged interfaces contain chains of multiple lengths.

The mechanical strength of a polymeric interface is achieved by molecular mechanisms that transfer stress across the boundary between the films. Numerous studies have shown that entanglements and friction between chains from opposite sides of the interface are among the major factors that control interfacial strength.
 Entanglements \cite{degennes71,doi88} are topological constraints that arise because polymers are long and strong covalent bonds prevent them from passing through each other.
Friction arises from van der Waals interactions between the polymers, which resist sliding of the molecules past each other \cite{schnell99,benkoski02}.
Both processes help anchor chains in opposing sides of the interface and impact the interfacial strength. These processes are expected to be strongly affected by the nature of the polymer including direct chemical interactions, total length of the molecules and their stiffness.
As noted above, the distribution of chain lengths may be different for welding
and healing, since fracture can produce a broad distribution of chain
lengths that impacts the healing process.
In this paper we contrast the evolution of strength during welding and healing
and examine the effect of chain stiffness on healing.

The significance of recovering bulk strength has led to a concerted effort to extract the factors that govern the strength of an interface.
Neutron reflectometry studies by Kunz and Stamm \cite{kunz96} followed the initial interdiffusion of polymers across an interface and distinguished different characteristic times for broadening of the interface.
Experimental studies by Schnell et al.~\cite{schnell98,schnell99} and Brown \cite{brown01} have shown that the mechanical tensile strength at polymer interfaces correlates very well with the interfacial width.
Interfacial width was measured by neutron reflectivity, while the interfacial strength was characterized by fracture toughness determined from a double cantilever beam test.
Schnell et al.~\cite{schnell99} showed that the rise of interfacial toughness under tension is related to the transition of the failure mechanism from simple chain pullout to crazing at the interface.
Recently, McGraw et al.~\cite{mcgraw13} have performed crazing measurements to probe the interdiffusion between two thin films of entangled polymers and showed that it takes less than one bulk reptation time to observe the structural characteristics of bulk crazes at the interface.
Similar conclusions about the recovery of interfacial strength have been
obtained from lap-joint shear tests \cite{kline88,parsons98,akabori06,boiko12}.
These studies determined the ultimate shear stress at failure and correlated
it with interdiffusion time.

\begin{figure}[htb]
\includegraphics[width=0.5\textwidth]{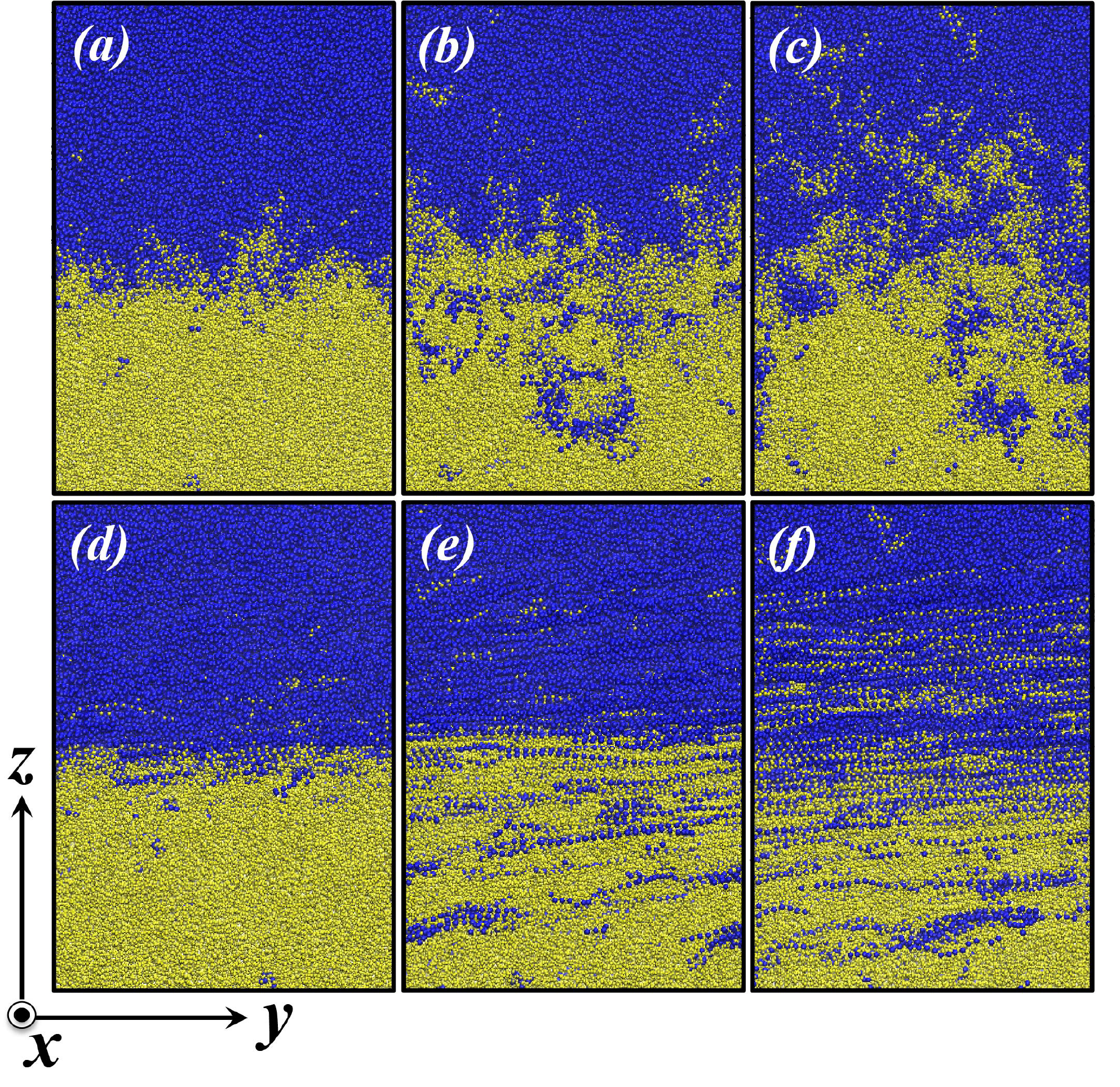}
\caption{Snapshots showing the evolution of the interface due to interdiffusion and shear after bonds crossing the plane $z=0$ are cut. Beads are colored based on their positions before the cut:  $z>0$ (blue) and $z<0$ (yellow). For clarity only a portion of the sample, $40a$ by $60a$ in the $y-z$ plane and $10a$ deep along $x$ is shown. Snapshots (a), (b) and (c) depict the interface at interdiffusion time $t=0.01M\tau$, $0.5M\tau$ and $7M\tau$ respectively. Snapshots (d), (e) and (f) show the corresponding states after a large shear strain $\gamma=12$ along $y$. Chains have initial length $N=500$ and $T=0.2 u_0/k_B$.}
\label{fig:snapshot}
\end{figure}

Theoretical and computational studies have further probed the dynamics of polymers at interfaces, providing detailed insight into the location and conformation of macromolecules. For miscible polymers, reptation dynamics, which was originally introduced by de Gennes \cite{degennes71} and later refined by Doi and Edwards \cite{doi88} to explain the dynamics of entangled chains in bulk melt, was able to describe \cite{jud81,prager81,prager83,kim83,adolf85,mikos89,wool95} diffusion across a polymer interface in a welding process. For immiscible polymers in which interpenetration of the films is limited, Helfand and Tagami \cite{helfand71,helfand72} calculated the equilibrium interfacial density profile using mean field theory.

Computer \cite {deutsch91, haire01, anderson04, pierce11, ge13, ge13b} simulations have been performed to study the interdiffusion across both miscible and immiscible polymer joints. For miscible polymers, results of Monte Carlo simulations \cite{haire01} of a lattice polymer model showed that the mass uptake between polymers agrees with the prediction of reptation dynamics. However, our recent molecular dynamics simulations \cite{pierce11,ge13} of thermal welding of two thin films of identical monodispersed polymers showed that the interfacial dynamics is controlled by the motion of chain ends, and is faster than predicted by reptation dynamics. For immiscible polymers, both Monte Carlo \cite{deutsch91,anderson04,ge13b} and molecular dynamics simulations \cite{ge13b} have found that the immiscibility arrests the interdiffusion and limits the equilibrium interfacial width.  

Recently, we have probed the interfacial strength of polymer interfaces in our molecular dynamics simulations \cite{ge13, ge13b} using a shear test that is similar to lap-joint shear experiments \cite{kline88,parsons98,akabori06,boiko12}. For miscible polymers, the interfacial strength increases with interdiffusion time, and saturates at the bulk strength. The dominant failure mode changes from pure chain pullout at the interface for short welding times to chain scission for long welding times, as in bulk failure. In contrast, even for weakly immiscible polymers, the narrow interface is unable to transfer stress upon deformation as effectively as the bulk polymer, and chain pullout at the interface is the dominant failure mode. These studies have shown that, as expected, entanglements play a critical role in resisting chain pullout and enhancing the interfacial strength. 

 In this work, we probe the interface between two polymeric films that were cut and allowed to heal as illustrated in Fig.~1.
 In contrast to thermal welding of separate parts of identical polymers \cite{jud81,wool81, prager81,prager83,kline88,mikos89,wool95,parsons98,akabori06,boiko12,wool08,blaiszik10}, scission  of polymers  results in formation of interfaces that consist of chains of multiple lengths. This polydispersity is key to the dynamics of healing of polymer films.  
Simulation results for healing are compared with previous findings for thermal welding. The strength of a healed interface formed by polymers with different stiffness is probed, and related to the molecular mechanisms of deformation and failure. These studies are correlated with the local density of entanglements extracted from primitive path analysis, a recent methodology  developed to enable one to identify and track the evolution of entanglements\cite{everaers04,kroger05,tzoumanekas06}. These simulations allow a direct measurement of various parameters that characterize the interfacial structure and reveal correlations between them and the interfacial strength.

Sec.~\ref{sec:method} presents the simulation model and methodology used in this study. Then Secs.~\ref{sec:dynamics} and \ref{sec:structure} present results for the diffusion dynamics and the evolution of molecular and entanglement structure at the interface during healing. An analysis of the evolution of interfacial strength is presented in Sec.~\ref{sec:strength}, highlighting the effects of polydispersity and chain stiffness. Finally a summary and conclusions are given in Sec.~\ref{sec:sum}. 

\section{\label{sec:method} Model and Methodology}

\noindent {\bf A. Simulation Model}
\medskip

Here we employ a coarse-grained bead-spring model \cite{kremer90} that captures well the properties of linear homopolymers.
Each polymer chain initially contains $N=500$ spherical beads of mass $m$.
Beads interact via the truncated and shifted Lennard-Jones potential
\begin{equation}
U_{\rm LJ} (r)=4u_0 [(a/r)^{12}-(a/r)^6-
(a/r_{\rm c} )^{12}+(a/r_{\rm c})^6] \ \ ,
\end{equation}
where $r$ is the distance between beads, $r_{\rm c}$ is the cutoff radius and $U_{\rm LJ} (r)=0$ for $r>r_{\rm c}$.
All quantities are expressed in terms of the molecular diameter $a$,
the binding energy $u_0$, and the characteristic time $\tau=a(m/u_0 )^{1/2}$.
To provide a rough mapping to real hydrocarbons we take
$a \sim 0.5$nm, which is a typical inter-polymer distance, and a binding
energy of order the glass transition temperature, $u_0 \sim 40$meV.
One then finds the units of force and stress are
$u_0/a \sim 13$pN and $u_0/a^3 \sim 50$MPa, respectively.
The latter is of the same order as yield stresses in glassy polymers.

For equilibration and healing runs we used the unbreakable finitely extensible nonlinear elastic (FENE) potential \cite{kremer90} to bond adjacent monomers along the backbone,
\begin{equation}
U_{\rm FENE} (r)=-\frac{1}{2}kR_0^2 \ln [1-(r/R_0 )^2] \ \ ,
\end{equation}
with $R_0=1.5a$ and $k=30 u_0 a^{-2}$ \cite{kremer90}.
For mechanical tests, chain scission plays an essential role and a 
simple quartic potential was used:
\begin{equation}
U_{Q} (r)=K(r-R_{\rm c} )^2 (r-R_{\rm c} )(r-R_{\rm c}-B)+U_0 \ \ ,
\end{equation}
with $K=2351u_0/k_B$, $B=-0.7425a$, $R_{\rm c}=1.5a$, and $U_0=92.74467u_0$.
This potential provides the same equilibrium bond length as $U_{\rm FENE}$ and prevents chains from crossing.
The bonds break at a force $f_b =240 u_0/a$ that is $\sim100$ times higher than the maximum attractive force for the interchain $U_{\rm LJ}$.
This force ratio has been used in previous simulations \cite{rottler02a,sides01,stevens01b,stevens01}, and is supported by typical experimental values \cite{odell86,creton92}. 
From the mapping to hydrocarbons in the previous paragraph, one finds
$f_b=240 u_0/a \sim 3$nN.
This is comparable to experimental estimates \cite{odell86,creton92}
and density functional theory results ($\sim 6$nN) for the force needed to break pentane backbone bonds \cite{foot0}.
Simulations with a range of potentials show that the breaking force is the most important aspect of the potential and its influence is discussed briefly in Sec.~IIIC.

An additional bond bending potential 
\begin{equation}
U_{\rm B} (\theta)=k_{\rm bend} (1+\cos\theta) \ \ ,
\end{equation}
is used to vary chain stiffness, where $\theta$ is the angle between two consecutive bonds along the chain.
Varying chain stiffness serves as a means to change the entanglement spacing.
As $k_{\rm bend}$ increases, the entanglement length $N_e$ decreases and polymer chains become more entangled.
Previous studies \cite{puetz00,hoy09,hou10} find that for $k_{\rm bend}/u_0 = 0$, $0.75$ and $1.5$ the corresponding $N_{\rm e}$ is approximately $85$, $39$ and $26$, respectively. 
Experiments and simulations\cite{wool95,rottler03} show that the mechanical response becomes independent of chain length for $N/N_e$ larger than 6 to 12.
We choose $N=500$ to be in this limit for all $k_{\rm bend}$.

All simulations were carried out using the LAMMPS parallel MD code \cite{plimpton95}.
The equations of motion were integrated using a velocity-Verlet algorithm.
For equilibration and healing runs, the time step $\delta t=0.01\tau$, while for shear strength tests the time step was reduced to $\delta t=0.005\tau$.
The temperature $T$ was held constant for the equilibration and healing runs by a Langevin thermostat 
with a damping constant $\Gamma=0.1\tau^{-1}$ \cite{kremer90}.
For the shear runs $\Gamma = 1 \tau^{-1}$, and was applied only in the $x-$ direction to avoid biasing the flow. A million $\tau$ will be abbreviated as $1 M\tau $.

\medskip
\noindent{\bf B. Bulk Simulations}
\medskip

To identify the relevant time and length scales for chain diffusion, we performed simulations of self-diffusion in bulk samples.
Three samples consisting of $M=500$ chains of length $N=500$ and $k_{\rm bend}/u_0 =0$, $0.75$ and $1.5$ respectively were constructed using the standard methodology discussed by Auhl et al.~\cite{auhl03}.
The unit cell was cubic and periodic boundary conditions were applied in all directions.
Samples were then equilibrated at pressure $P=0$ and $T = 1.0 u_0/k_B$ in an NPT simulation with $r_c=2.5 a$.
The root mean square radius of gyration $<R_g^2>^{1/2}/a= 11.9\pm 0.1$, $13.1\pm 0.1$, and $15.1\pm 0.1$ for $k_{\rm bend}/u_0=0$, $0.75$ and $1.5$ respectively.
The mean squared end-to-end distance $<R^2>=6.0\pm 0.02<R^2_g>$ in all three cases, as expected for Gaussian chains.

\begin{figure}[htb]
\includegraphics[width=0.45\textwidth]{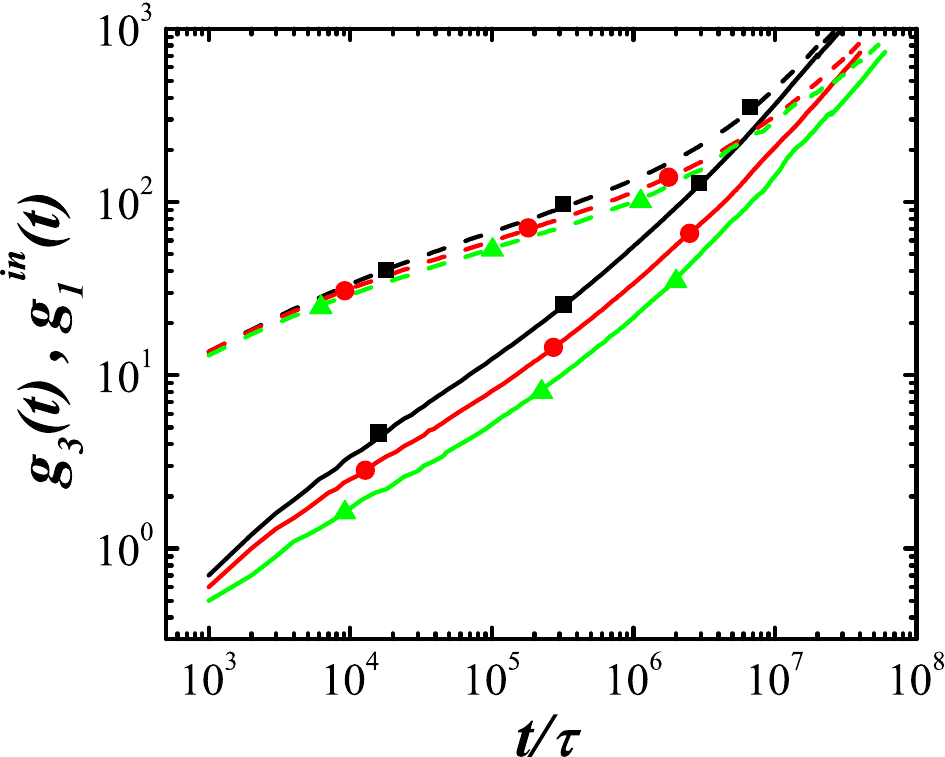}
\caption{
Mean squared displacement for the center of mass $g_3(t)$ (solid lines) and the center 10 individual beads of the chain $g_1^{in}(t)$ (dashed lines) during self-diffusion in bulk systems with $N=500$ and $k_{\rm bend}/u_0 =0$ (black squares), $0.75$ (red circles) and $1.5$ (green triangles).}
\label{fig:msd}
\end{figure}

After equilibration, simulations were run for $80M\tau$ for $k_{\rm bend}/u_0=1.5$ and $50M\tau$ for $k_{\rm bend}/u_0=0.75$ and $0$.
Results for the mean squared displacement for the chain center of mass (CM) $g_3(t)=\left<{(\mathbf{r}_{CM}(t)-\mathbf{r}_{CM}(0))}^2\right>$, and the center 10 individual beads of the chain $g_1^{in}(t)=\left<{(\mathbf{r}_{i}(t)-\mathbf{r}_{i}(0))}^2\right>$ are shown in Fig.~\ref{fig:msd}.
Based on the tube model \cite{doi88,puetz00}, the increase of $g_1^{in}(t)$ should cross over from $t^{1/2}$ to $t^{1/4}$ scaling, and that of $g_3(t)$ from $t$ to $t^{1/2}$, when entanglements set in at the entanglement time $\tau_e$.
From Fig.~\ref{fig:msd} we estimate that
$\tau_e = 10^4\tau$, $6 \cdot 10^3\tau$ and $4 \cdot 10^3\tau$ for
$k_{\rm bend}=0$, $0.75$ and $1.5$, respectively.
As expected, the decrease in $N_e$ with increasing chain stiffness lowers $\tau_e$.

Stronger entanglement effects associated with smaller $N_e$ prolong the disentanglement time $\tau_d$ that is required for the entire chain to diffuse by a distance on the order of $<R^2>^{1/2}$ and get disentangled from the initial confining chains at $t=0$.
At the disentanglement time there is a crossover to the diffusive regime where both $g_1^{in}(t)$ and $g_3(t)$ increase linearly with time.
For $k_{\rm bend}/u_0 =0$, the two curves in Fig.~\ref{fig:msd} overlap near $\tau_d \sim30M \tau$ and the mean-squared displacement
reaches $<R^2>$ at about $25 M\tau$.
For $k_{\rm bend}/u_0 =0.75$ and $1.5$, $\tau_d$ is larger than the times which are presently accessible by simulation.
Extrapolating to the points where the curves would reach $<R^2>$ gives values of $\tau_d$ that are of order
$60M\tau$ and $120M\tau$ for $k_{\rm bend} =0.75$ and $1.5$, respectively.

\medskip
\noindent{\bf C. Interface Simulations}
\medskip

For the healing studies, we constructed three films of $M=9600$ chains of length $N=500$ beads or a total of $4.8$ million beads with $k_{\rm bend}/u_0 =0$, $0.75$ and $1.5$.
Periodic boundary conditions were applied along the $x$- and $y$- directions with dimensions $L_x=700a$ and $L_y=40a$.
The thickness in the $z$-direction was maintained at $L_z=200a$ using two repulsive confining walls.
Snapshots in Fig.~\ref{fig:snapshot} illustrate part of the cross section of such a thin film in the $yz$-plane.
After construction, all films were well equilibrated at a temperature $T=1.0 u_0/k_B$ and pressure $P=0$ with $r_{\rm c}=2.5a$, by allowing expansion/contraction along the $x$-direction.
The total change in $L_x$ was minimal ($\sim 2\%$) and the final
monomer density of all systems was $\rho a^3=0.89$.

A fracture plane was introduced by cutting almost all bonds that crossed the middle plane at $z=0$.
An exception was made when two crossing bonds were separated by less than three monomers.
In this case only one of the bonds was cut to minimize the number of extremely short chains.
A small number of bonds with length 3 and below are formed from chains that cross the mid-plane near a chain end.
Both the bonding potential and any bond bending terms associated with cut bonds were removed
from the Hamiltonian.

\begin{figure}[htb]
\includegraphics[width=0.4 \textwidth]{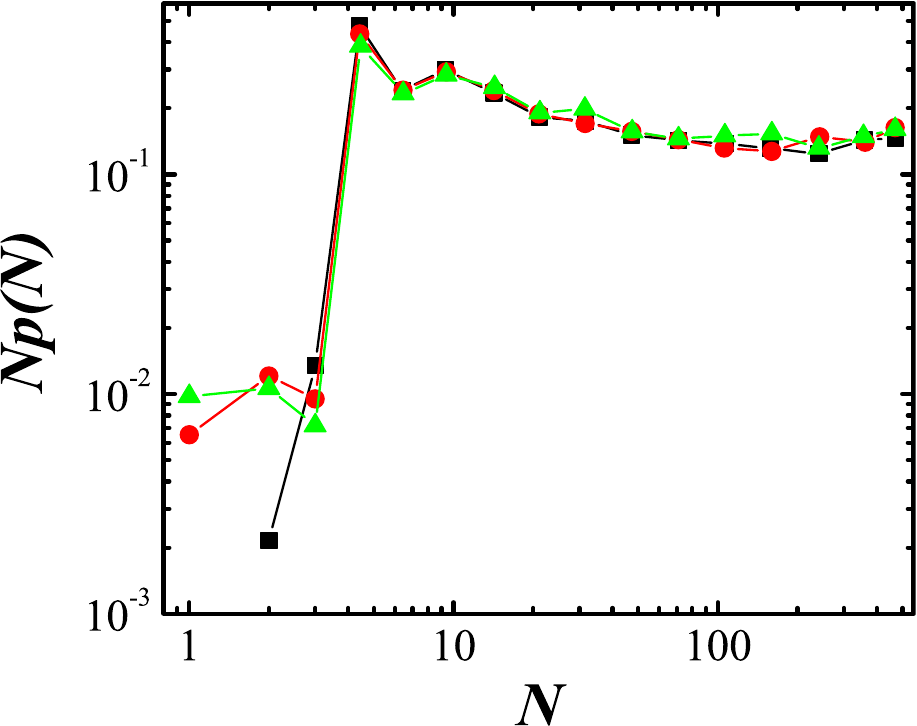}
\caption{The probability $N p(N)$ that a monomer is in a cut segment of length $N$
for $k_{\rm bend}/u_0 =0$ (black squares), $0.75$ (red circles) and $1.5$ (green triangles).
Uncut chains are not included in the statistics.
}
\label{fig:polydisperse}
\end{figure}

Figure \ref{fig:polydisperse} shows the resulting distribution of segment lengths from cut chains.
The probability $p(N)$ that a segment has length $N$ scales roughly as $1/N$.
This means that the probability $N p(N)$ of a monomer being in a chain of length $N$
is nearly independent of $N$.
The probability distribution is nearly independent of chain stiffness, but there is 
an increase in the total number of cut segments with increasing stiffness
(8691, 9507 and 10047 for $k_{\rm bend}/u_0 = 0$, $0.75$ and $1.5$, respectively).
Stiffer chains are less likely to bend back across the interface, so more chains are cut into fewer pieces.
Since $n$ cuts produce $n+1$ segments, the same number of cut bonds leads to more segments.
The suppression of cut segments of length less than 3 also has less impact on stiffer chains.

After bond cutting, the systems were held at $k_B T = 1.0 u_0$ and allowed to
interdiffuse.
Configurations for a series of healing times
were then
quenched rapidly below the glass temperature $T_g \approx 0.35 u_0/k_B$.
First the cutoff radius was reduced to $r_{\rm c}=1.5a$, in order to facilitate comparison with past mechanical studies \cite{rottler02a, rottler02b, rottler03, hoy07, hoy08} and reduce density changes.
Then the temperature was quenched at constant volume with a rate $\dot{T} =-10^{-3}  u_0/(k_B \tau)$ to $T=0.5 u_0/k_B$ where $P=0$.
Subsequent quenching to $T=0.2 u_0/k_B$ was done at $\dot{T} = -2 \times 10^{-4} u_0/(k_B \tau)$ and $P=0$.
A Nose-Hoover barostat with time constant $50 \tau$ was applied to $P_{\rm xx}$ and $P_{\rm yy}$.
This quench protocol is rapid enough to preserve the interfacial structure produced by
interdiffusion in the melt state at the given healing time.
We verified that our conclusions were not sensitive to the details of the quench protocol or geometry.
This is consistent with past experiments and simulations
that find that the quench rate and aging affect the initial yield stress \cite{rottler05} but not
strain hardening and failure \cite{hasan93, hoy08}.
Snapshots in the top row of Fig.~\ref{fig:snapshot} depict an interface during healing at three interdiffusion times after being quenched into the glassy state.

As in previous simulations \cite{ge13},
we determined the interfacial strength using a shear test
that is similar to a lap-joint shear experiment \cite{kline88,parsons98,akabori06,boiko12}.
To focus on a region of width $H$ near the interface,
atoms in top ($z > H/2$) and bottom ($z< -H/2$) layers were held rigid
and displaced at constant velocity in opposite directions in the $xy-$plane.
In most cases $H=50 a$ and studies with double and quadruple this value produced little change in the maximum shear stress compared to statistical fluctuations ($< 5$\%).
The shear direction had a bigger effect.

Results are presented for shear along the $y$-axis ($L_y=40a$) because they were more reproducible.
The rigid top and bottom boundaries screen elastic interactions between regions separated by more than $H$.
When the system is sheared along the $y$-axis these different regions are effectively independent.
Averaging over the system is equivalent to averaging over independent smaller simulations.
To test this we calculated the local stress on regions of length $100a$ along the $x$ axis.
The deviations from the mean stress in each region are uncorrelated.
This confirms that our simulations are equivalent to several independent realizations and allows us to estimate the statistical error in the average stress.
The error in the peak stress is $\sim 2\%$, while the error is $\sim 4\%$ at early stages of the shear test where the stress is lower.
Similar estimates were obtained by comparing to simulations with $L_x=180a$ and $L_y =40a$. 
In contrast, when the system is sheared along the long $x-$ axis ($L_x=700a$) different
regions are coupled because shear causes them to pass over each other.
When the interface achieves bulk strength, the initial failure occurs at different heights in different regions.
If the sample is sheared along the long $x-$ axis ($L_x=700a$), these failure zones must be connected by diagonal shear planes and the peak stress varies by $20$\% with small changes in initial conditions.
Experiments may involve much more complicated dissipation mechanisms and loading conditions, particularly under tensile testing.

The velocities of top and bottom rigid layers are chosen to produce an average strain rate in the film,
$d\gamma/dt=2 \times 10^{-4} \tau^{-1}$, that is low enough that it does not affect the mode of failure and stress has time to equilibrate across the system \cite{rottler03c}.
The shear stress $\sigma$ is determined from the mean lateral force per unit area applied by the top and bottom layers and is consistent with the time average of the microscopic stress in the sample.
All of the shear tests are run at a temperature $T= 0.2 u_0/k_B$ to a large shear strain $\gamma \sim 12$.
Snapshots in the bottom row of Fig.~\ref{fig:snapshot} illustrate the interfacial structure at large $\gamma$ for different interdiffusion times.  

Two additional geometries were studied to help calibrate and interpret the healing data.
To determine the ideal strength of bulk samples we used systems with the same geometry and interactions but without cutting bonds.
To compare healing of a fracture plane with welding of separate polymer parts, we also simulated welding of two polymer thin films using the same model \cite{ge13}.
Each film contained $M=4800$ chains of length $N=500$ beads with $k_{\rm bend}=0$.
Dimensions in the $xy$-plane were identical to those in the healing samples.
The two films were well equilibrated, brought into close contact at the welding plane $z=0$,
and allowed to interdiffuse in the $z-$direction at $T=1.0 u_0/k_B$.
Simulation protocols for interdiffusion, quenching and the subsequent mechanical test were all similar to those used for healing. More details can be found in our previous paper \cite{ge13}. 

\begin{figure}[htb]
\includegraphics[width=0.48\textwidth]{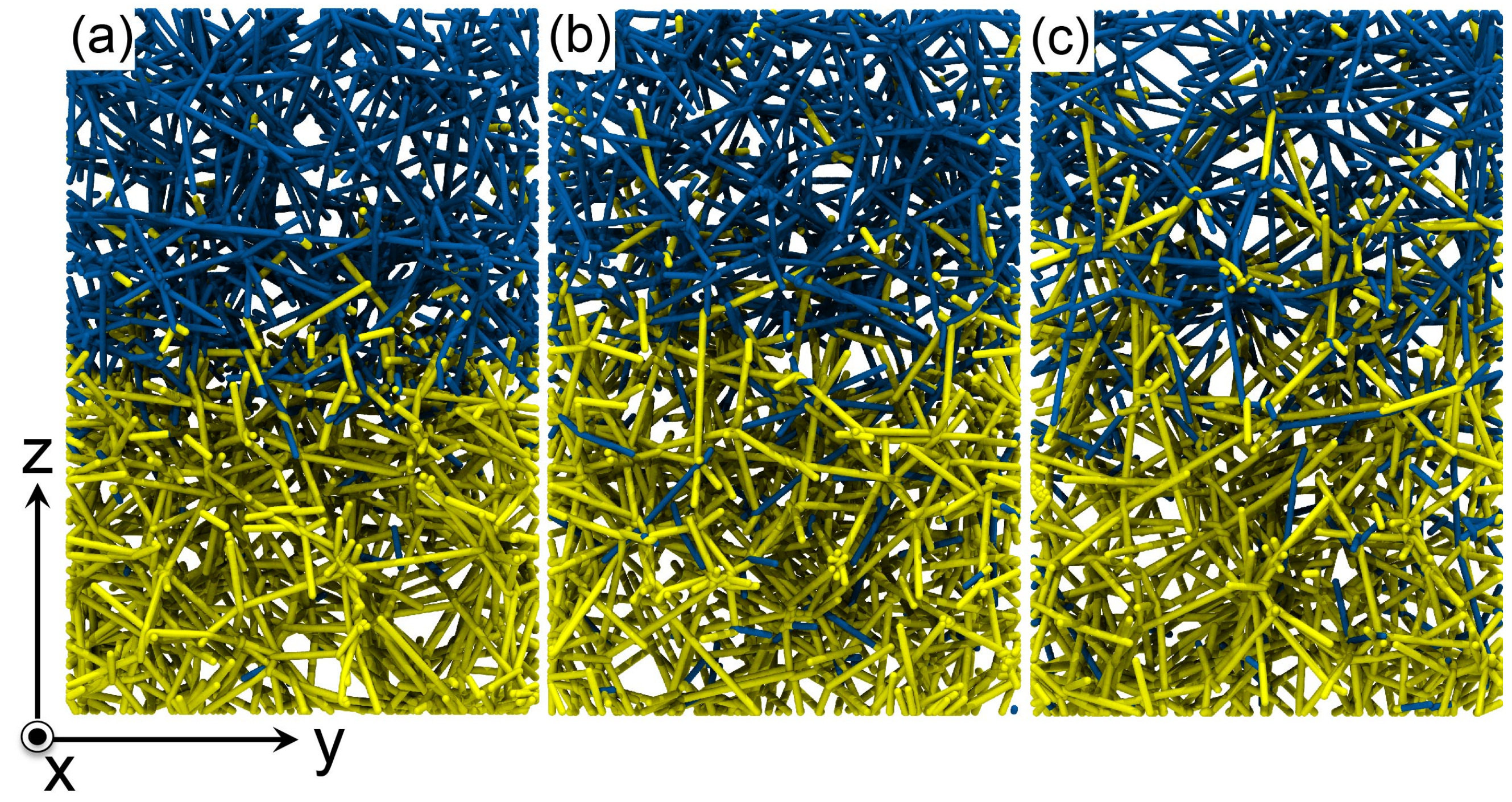}
\caption{Snapshots of the primitive paths for the states in Fig.~\ref{fig:snapshot}(a), (b) and (c). Beads are colored based on their positions before cutting: $z>0$ (blue) and $z<0$ (yellow).}
\label{fig:PPA}
\end{figure}

Entanglements in all states of different samples are identified using the Primitive Path Analysis (PPA) algorithm \cite{everaers04}.
Chain ends are frozen and tensile forces are introduced to shrink the contour length without allowing chain crossing.
To limit excluded volume effects, the chain diameter is then reduced by a factor of 4 and the contour is minimized again \cite{hoy07b}.
The resulting configuration is a network of primitive paths. Figure~\ref{fig:PPA} shows the primitive paths in a healing interface at different interdiffusion times.
We identify the contacts between primitive paths as topological constraints (TCs) associated with entanglements.
As in previous studies \cite{hoy07b,tzoumanekas06,everaers12}, the resulting density of TCs, $\rho_{TC}$, is proportional to the density of
entanglements, and variations in the density are not sensitive to the precise details of the identification procedure.   

\section{Results}
\label{sec:result}

\subsection{Interfacial Dynamics}
\label{sec:dynamics}

\begin{figure}[!htbp]
\centerline{\includegraphics[width=0.4\textwidth]{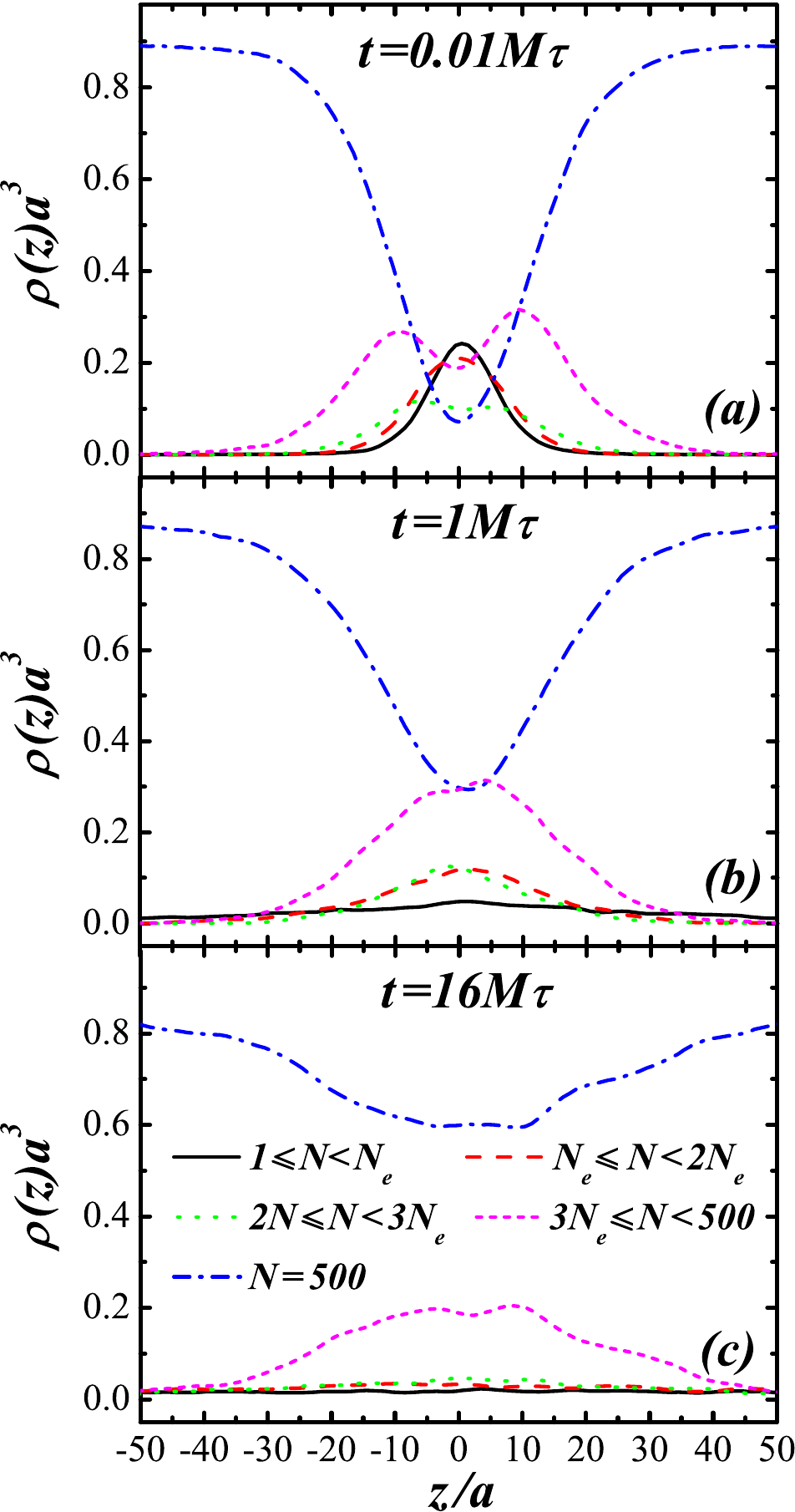}}
\caption{Density profiles $\rho(z)$ of monomers belonging to segments of the indicated length
for fully flexible chains with $k_{\rm bend}=0$ at (a) $t=0.01M\tau$, (b) $1M\tau$ and (c) $16M\tau$.
There are roughly equal numbers of monomers in segments of all length (Fig.~\ref{fig:polydisperse}), so the total number of monomers
with $3N_e < N < 500$ is roughly 3 times larger than the number in the other length ranges which have width $N_e$.}
\label{fig:density}  
\end{figure}

Cutting bonds at $z=0$ introduces polydispersity in chain length near the interface that has strong consequences for both the shear strength and the dynamics of healing.
Figure~\ref{fig:density}(a) shows the distribution of monomers from chains of different length shortly after cutting the $k_{\rm bend}=0$ system.
The probability of being on an uncut chain is strongly suppressed near $z=0$, while most short segments are close to $z=0$.
The distribution of intermediate length segments has a double peak structure.
While both ends of these segments are at the interface, the remainder is all on one side or the other.
The longer the segment, the farther the center of mass is displaced from the mid-plane.

The self-diffusion constant $D$ scales with $N$ as $N^{-1}$ for short chains,
but drops as $N^{-2}$ for longer chains\cite{doi88}.
The crossover between these two regimes occurs near $N_e$ when entanglements start to affect the dynamics.
These variations in $D$ lead to striking differences in the redistribution of segments during healing.
By $1M\tau$ (Fig.~\ref{fig:density}(b)), segments with $N < N_e$ have spread nearly uniformly throughout our finite system.
For longer segments, the two initial peaks in $\rho(z)$ have merged into a single peak,
whose width ($\sim 35 a$) seems relatively insensitive to chain length.
At the longest time, $16M\tau$, segments with $N < 3N_e$ have diffused away from the interface.
Longer chains remain at the interface, but are spread over a slightly wider range $\sim 50a$.
As discussed below,
the slow diffusion of these cut segments is important in the final approach to bulk strength.

\begin{figure}[!htbp]
\centerline{\includegraphics[width=0.4\textwidth]{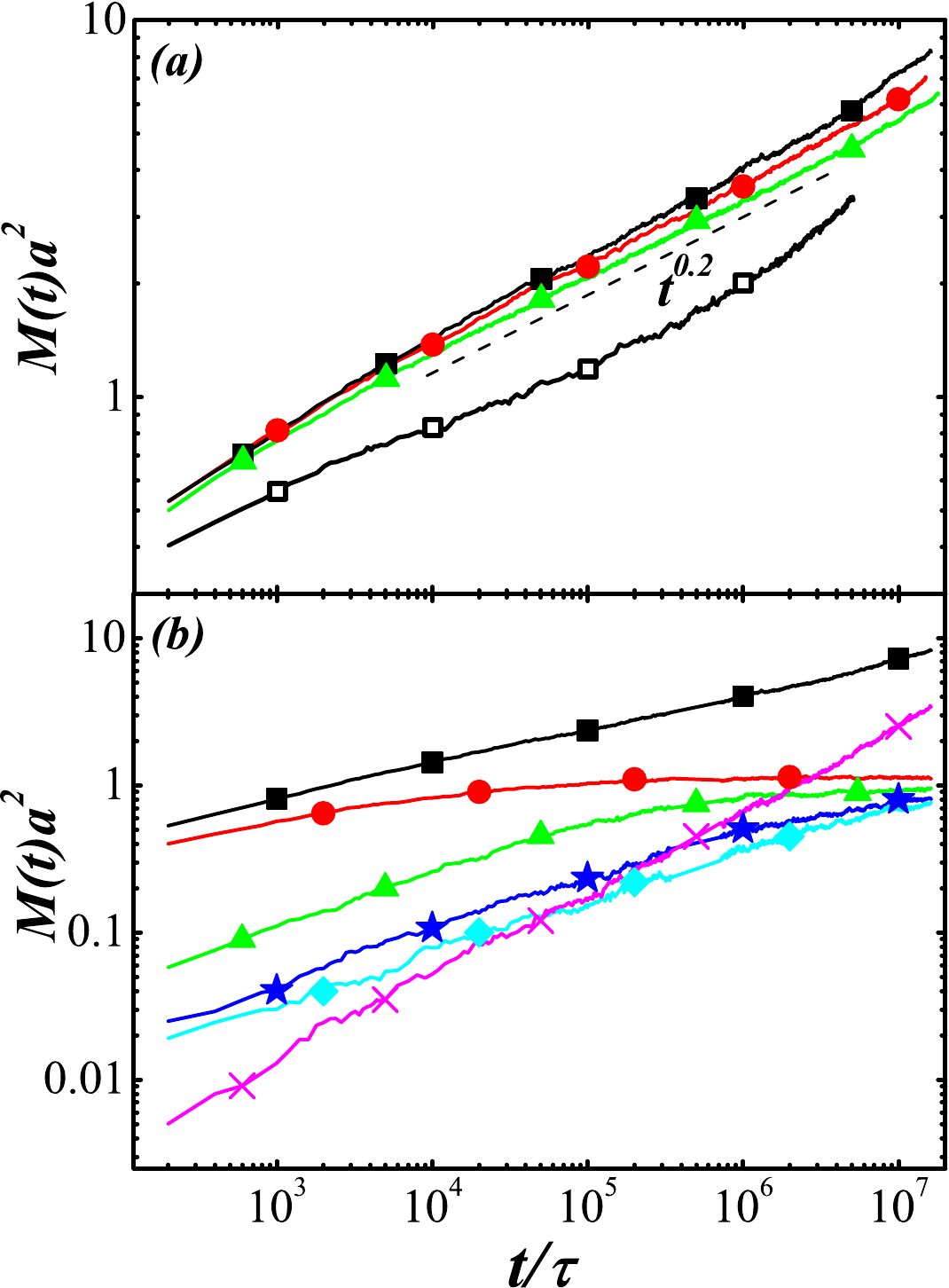}}
\caption{(a) Mass uptake per unit area $M(t)$ as a function of interdiffusion time $t$ for healing samples with $k_{\rm bend}/u_0=0$ (black squares), $0.75$ (red circles) and $1.5$ (green triangles).
A dashed line shows the $t^{0.2}$ scaling that curves follow at $t > t_e$.
Also shown is $M(t)$ for a welding sample with $k_{\rm bend}/u_0=0$ (open black squares). 
(b) Mass uptake for all monomers (black squares) and monomers in segments with $1\leq N<N_e$ (red circles), $N_e\leq N<2N_e$ (green triangles), $2N_e\leq N<3N_e$ (blue stars), $3N_e\leq N < 4N_e$ (cyan diamonds) and uncut chains (magenta crosses) during healing of a $k_{\rm bend}=0$ sample. 
}
\label{fig:mass}
\end{figure}

Healing occurs by the diffusion of chains across the interface to create interfacial entanglements \cite{wool95}.
A common experimental measure is the mass uptake, which corresponds to
the number of beads that have crossed to the opposite side of the initial interface.  
Figure \ref{fig:mass}(a) shows mass uptake per unit area, $M(t)$,
as a function of time for healing samples with different stiffness
and a welding sample with flexible chains.
Increasing the chain stiffness lowers the rate of uptake.
This is consistent with the slower diffusion of stiffer chains in
Fig.~\ref{fig:msd} and each curve shows a small change
in slope at a time that is comparable to the corresponding bulk $\tau_e$.
At longer times the healing curves all show an extended power law regime $M(t) \sim t ^\alpha$ with exponent $\alpha \approx 0.2$.

Theories for monodisperse chains predict that at short time the mass uptake should follow a power law consistent with Rouse dynamics in a tube, $M(t) \sim t^\alpha$ with $\alpha=1/4$ \cite{jud81,prager81,prager83,kim83,adolf85,mikos89,wool95}. The uptake should then cross over to the value of $\alpha=1/8$ expected for reptation.
All data for both healing and welding in Fig.~\ref{fig:mass} show a much faster uptake than expected from reptation.
The results for 
welding are consistent with those obtained by Pierce {\it et al.} \cite{pierce11}.
They associated the faster interdiffusion with faster motion of chain ends, which have a higher concentration near the welding interface.
Here we find that diffusion of chains across the interface is significantly faster for healing than welding.
Fracture produces a high density of short chains at the interface that diffuse rapidly
(Fig. \ref{fig:density}).

The importance of polydispersity is illustrated in
Fig.~\ref{fig:mass}(b), which shows
the mass uptake associated with segments of different length
for healing of flexible chains ($k_{\rm bend}=0$).
Unentangled segments ($N < N_e$) dominate the uptake at short times and
progressively longer chains become important at later times.
For each length the uptake saturates at long times when the chains
have moved by a distance comparable to
their initial distance from the interface, which is of order their radius of gyration.
The curves saturate at the same uptake because they contain the same range of lengths and there are equal numbers of monomers in cut chains of each length (Fig. \ref{fig:polydisperse}).

For each range of chain lengths the mass uptake rises with a power law higher than expected. For cut chains, $\alpha \sim 0.35$ instead of 1/4.
Over the entire time interval, the mass uptake of uncut chains increases as $t^\alpha$ with $\alpha = 0.56$.
These higher than expected exponents do not represent
superdiffusive motion of the chains in their tubes.
Instead they reflect the mimimum at $z=0$ in the local density of monomers in longer
chains immediately after cutting (Fig.~\ref{fig:density}).
Fewer monomers in longer chains are available
to diffuse across the interface at early times and the density of chains that can contribute to the mass uptake rises with time.

\begin{figure}[!htbp]
\centerline{\includegraphics[width=0.4\textwidth]{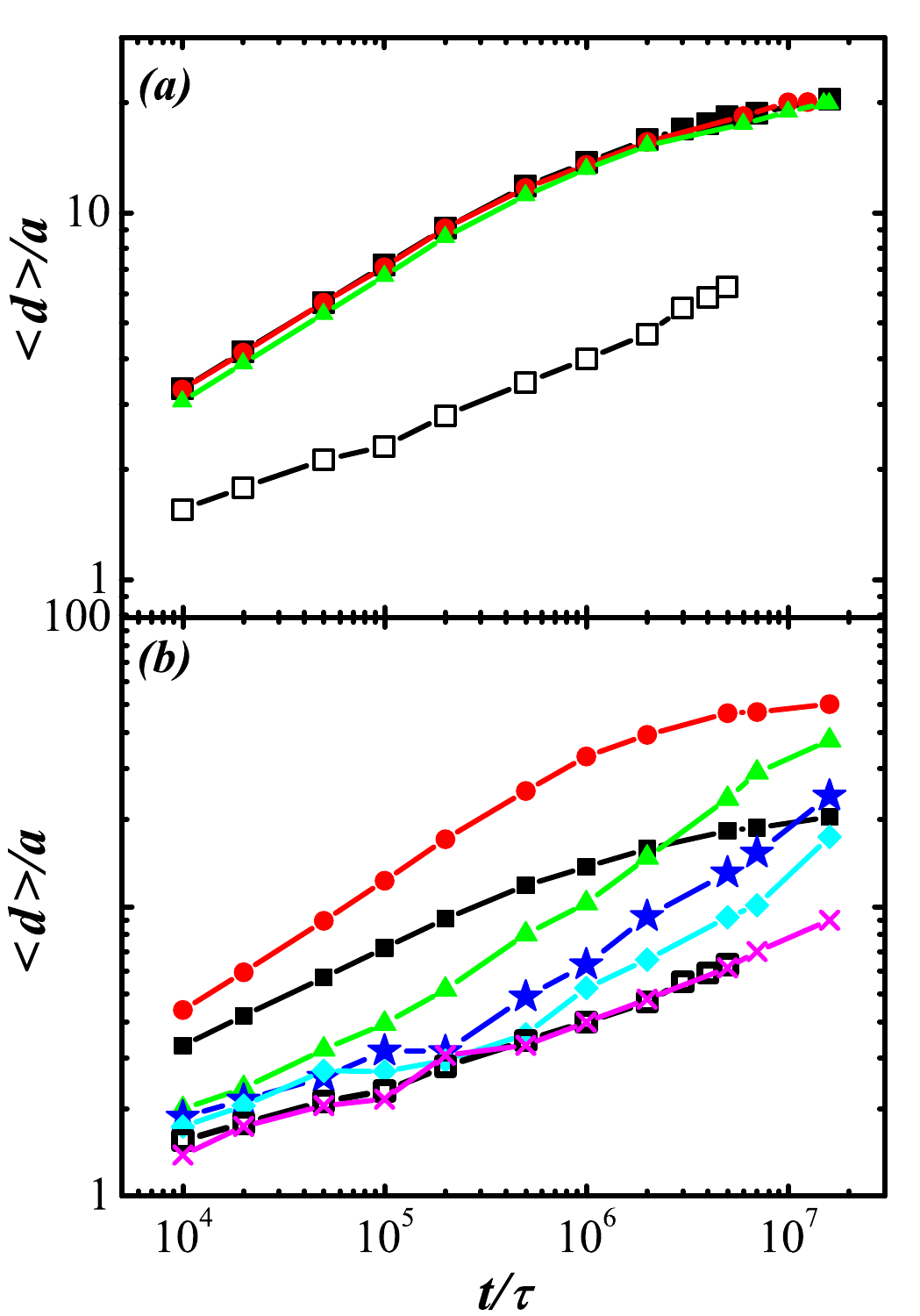}}
\caption{(a) Average interdiffusion depth $\left<d\right>$ of monomers from one side into the opposite side for the same systems as in Fig. \ref{fig:mass}.
(b) Separate averages for monomers in segments of the indicated lengths in healing
of $k_{\rm bend}=0$ chains and for all chains during welding (open squares).
For healing, separate results are shown for all monomers (black squares) and monomers belonging to chains with $N<N_e$ (red circles), $N_e\leq N<2N_e$ (green triangles), $2N_e\leq N<3N_e$ (blue stars), $3N_e\leq N <4N_e$ (cyan diamonds) and $N=500$ (magenta crosses).
}
\label{fig:depth}
\end{figure}

Another quantity of interest is how far monomers have moved across the interface, since a minimum distance is needed for entanglements to form \cite{ge13,ge13b,degennes89}.
Figure \ref{fig:depth}(a) shows the average interpenetration distance $\left< d \right>$ for all monomers in healing samples of different stiffness and for welding of flexible chains.
The rate of change decreases with increasing chain stiffness and is higher for healing than welding because of the polydispersity in cut samples.
Figure \ref{fig:depth}(b) shows the separate average interpenetration distances evaluated over
monomers from segments of the indicated range of lengths in the case $k_{\rm bend}=0$.
For short, unentangled chains, $\left< d\right>$ rises in a nearly Fickian manner, $\sim t^{0.5}$, and then saturates.
This saturation is a finite-size effect and the final value corresponds to monomers being evenly spread
across the sample as evidenced by Fig.~\ref{fig:density}.
In an infinite sample, these short segments would continue to leave the interfacial region. We show in Sec. \ref{sec:timestrength} that their concentration is so low that they do not have a significant impact on interfacial strength.

As chain length increases, the rise in $\left< d \right>$ approaches the $t^{1/4}$ time-dependence expected for reptation.
None of the entangled chains diffuse far enough for their motion to be limited by the finite system-size.
The behavior of uncut chains in healing and welding is nearly indistinguishable.
This is surprising in light of the very different rates of mass uptake for healing and welding (Fig.~\ref{fig:mass}(a)).
The suppression of $\rho(z)$ near $z=0$ (Fig.~\ref{fig:density}) means that fewer uncut chains
pass through the interface in healing than in welding.
However, Fig.~\ref{fig:depth}(b) shows that those that do pass through the interface travel about the same distance.

Zhang and Wool \cite{zhang89} calculated the concentration profile $\rho_1(z)$ for an interface between identical monodispersed polymers during welding for the minor chain model \cite{kim83}, and derived the time dependence of $\left<d\right>$ and $M(t)$ from the evolution of $\rho_1(z)$.
In particular, they found
$M(t) \sim t^{3/4}$ from a $t^{1/4}$ increase in the number of chains crossing the interface and an average contour length rising as $t^{1/2}$.
Later, Kim {\it et al.} \cite{kim96} used the same method to study the healing of fractured polymers by interdiffusion.
Compared with Zhang and Wool's results, they found that the increase of chain end density at the fracture plane leads to faster broadening of $\rho_1(z)$, and much larger $\left<d\right>$ and $M(t)$ for the same diffusion time.
Kim {\it et al.} also showed that the time dependence of $M(t)$ is changed from $t^{3/4}$ for welding to $t^{1/2}$ for fracture healing.
While our results do show faster mass uptake for healing, they are not consistent with the predicted power laws over the time regime shown in Fig.~\ref{fig:mass}.
As we discuss in the next section, one reason is that the number of interfacial loops remains nearly constant, rather than rising as $t^{1/4}$.

\subsection{Interfacial Structure}
\label{sec:structure}

Chains that diffuse across the interface do not necessarily contribute to the interfacial strength.
One idea is that interfacial strength is related to loops along chains
which extend into the opposing side and return.
The friction needed to pull these loops out of the opposing side will resist interfacial failure
and long loops will become entangled and provide even greater strength.
We identified interfacial loops by searching along each chain for segments that cross $z=0$ and return.
Because thermal fluctuations lead to a blurring of the interface, only loops of length $l> 3$ monomers are included in the analysis.

\begin{figure}[!htbp]
\centerline{\includegraphics[width=0.4\textwidth]{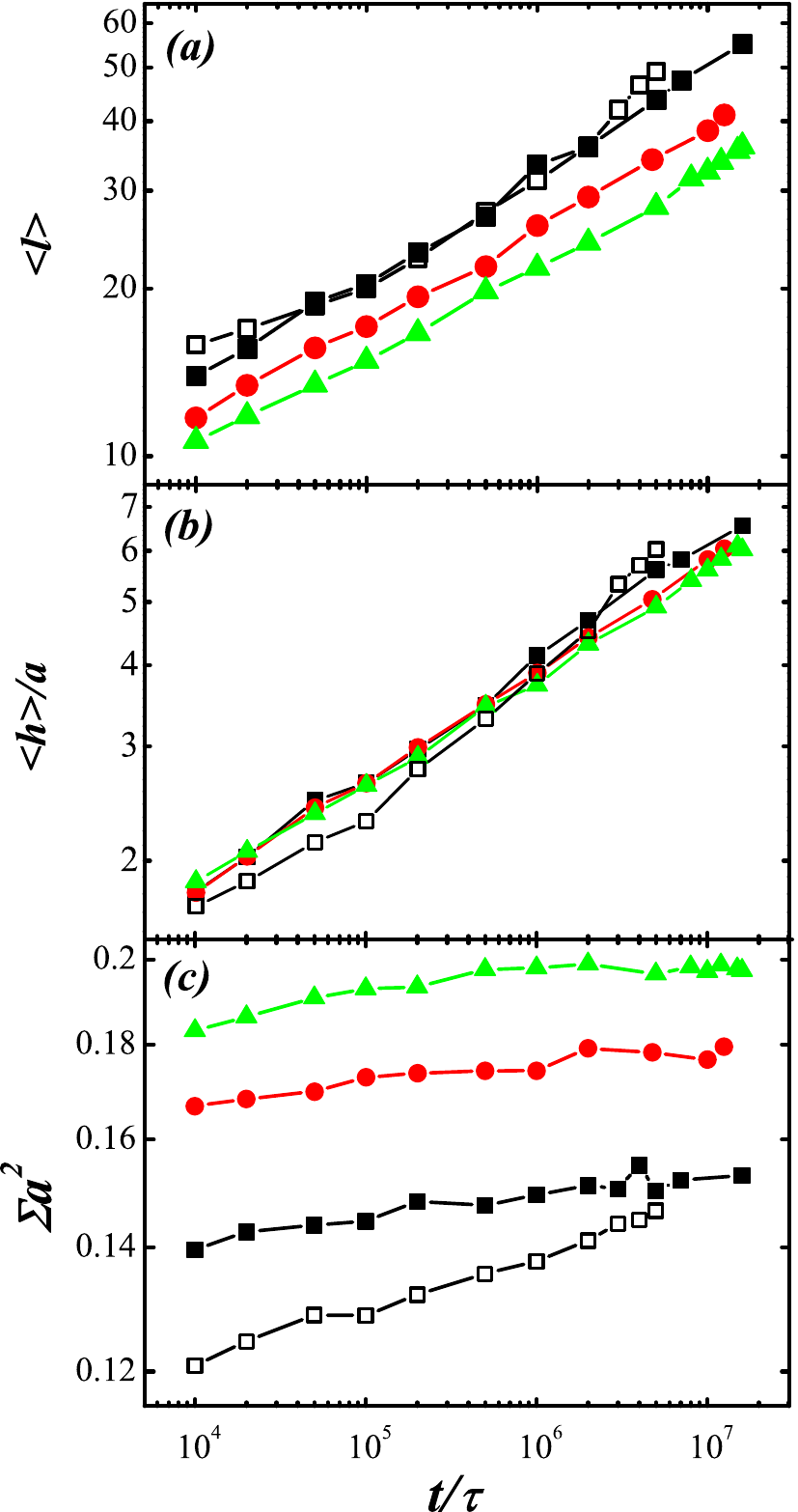}}
\caption{(a) Average contour length $\left<l\right>$ and (b) average depth $\left<h\right>$ of loops at the interface for the healing samples $k_{\rm bend}/u_0=0$ (black squares), $0.75$ (red circles) and $1.5$ (green triangles), and the welding sample with $k_{\rm bend}=0$ (open black squares).
(c) Time dependence of the areal density $\Sigma$ of interfacial loops contributing
to $\left< l \right>$ and $\left< h \right>$.
Only interfacial loops that have more than $3$ monomers are considered.
}
\label{fig:loops}
\end{figure}

The growth of the average length $\left<l\right>$ of interfacial loops with time $t$ for healing samples with different chain stiffness and welding of fully flexible chains is shown in Fig.~\ref{fig:loops}(a).
As before, increasing chain stiffness decreases diffusion and thus slows the growth of $\left< l \right>$.
More surprising is that the results for healing and welding samples with the same stiffness are almost the same.

Note that $\left< l \right>$ does not depend explicitly on the total number of loops, which should also influence strength.
Figure \ref{fig:loops}(c) shows the areal density of interfacial loops $\Sigma$ for the same systems.
Although $\left< l \right>$ is nearly the same for healing and welding of flexible chains, the areal density of loops is slightly lower at short times.
Increasing chain stiffness increases the density of interfacial loops.
In all cases, there is surprisingly little increase ($<20$\%) in the number of loops over 3 decades in time.
This is a striking contrast to the prediction of $t^{1/4}$ scaling by
Zhang and Wool \cite{zhang89}.

Loops that extend farther into the opposing surface may also provide greater strength.
Figure \ref{fig:loops}(b) shows the average depth $\left< h \right>$ of all monomers that are part
of interfacial loops.
Note that this is different from the interdiffusion depth $\left< d \right>$ plotted in Fig.~\ref{fig:depth}.
All chain segments contribute to $\left< d \right>$, while only loops with both ends at $z=0$ contribute to $\left< h \right>$.
At large $t$, this excludes unentangled or marginally entangled short chains that have diffused deep into the opposite side, and therefore $\left< h \right>$ could better reflect the interfacial structure related to shear strength.

The healing samples all have nearly the same $\left< h \right>$, even though $\left< l \right>$
decreases with stiffness. Assuming loops have a nearly equilibrium configuration,
$\left< h \right>$ should be of order the radius of gyration of segments of length $\left< l \right>$ and related to the positions of the bumps in $\rho(z)$ for intermediate chain lengths immediately
after cutting (Fig.~\ref{fig:density}).
The stiffer chains have a larger radius of gyration for the same
number of monomers which seems to compensate for the smaller $\left< l \right>$.

Several theoretical models relate the interfacial strength to the areal density of bridging loops across the interface \cite{prager81,adolf85,mikos88,mikos89,benkoski02, silvestri03}.
Some define bridging loops as chain segments intersecting both the plane at $z=\delta$ above and that at $z=-\delta$ below the interface ($z=0$), where $\delta$ is the end-end distance of the random walk between entanglements \cite{prager81,adolf85}.
Others define bridging loops as chain segments that connect two subsequent entanglements placed on different sides of the interface \cite{mikos88,mikos89,silvestri03}, or chain segments subject to at least one interfacial entanglement between opposite sides \cite{benkoski02}.
All of these assume that entanglements with both sides are required for effective stress transfer across the interface. From Fig.~\ref{fig:loops} one can determine when the mean loop length becomes comparable to $N_e$
or $\left< h \right>$ is comparable to $\delta$, and how rapidly the number of loops grows.
However the relation between these measures and entanglements is indirect. We therefore have also followed explicitly the development of entanglements.

\begin{figure}[!htbp]
\centerline{\includegraphics[width=0.45\textwidth]{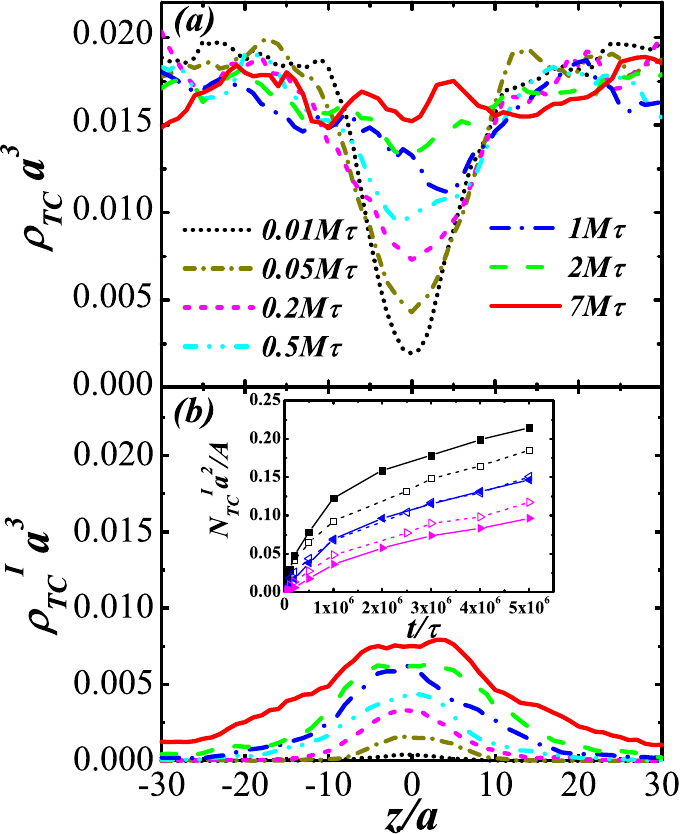}}
\caption{Density profiles at different interdiffusion times for TCs between chains from (a) either side and (b) opposite sides (interfacial TCs) of the interface in the healing sample of fully flexible chains ($k_{\rm bend}=0$). Only interior TCs with the number of monomers to the nearest chain end $n_{end}$ larger than $30$ are included. Inset of (b) shows the development of the areal density of interfacial TCs with interdiffusion time for both healing (filled symbols) and welding (open symbols) samples with $k_{\rm bend}=0$. Results are shown for TCs with $n_{end}>30$ (black squares), $50$ (blue left triangles) and $70$ (red right triangles) respectively.
}
\label{fig:TCs}
\end{figure}

The increase of entanglements across the interface during healing is tracked by applying the PPA at different interdiffusion times (Fig. \ref{fig:PPA}).
The density of topological constraints (TCs) between primitive paths provides a measure of the entanglement density. 
TCs that are near the end of a chain are not expected to contribute to mechanical strength because they can be removed by small displacements.
To account for this we remove TCs when the number of monomers $n_{end}$ to the nearest chain end is less than a threshold.
The density profiles of TCs between chains from any side and those between chains from opposite sides (interfacial TCs) are shown in Fig.~\ref{fig:TCs} for the healing sample of fully flexible chains.
Here only chains with $n_{end}>30$ are included, since the
distance between TCs along a chain is about 30 for $k_{bend}=0$.

At short times the density of entanglements is comparable to the bulk value far from the interface,
but reduced almost to zero near the interface ($|z|<10a$).
Similar behavior is observed for welding samples\cite{ge13,ge13b}, but they also exhibit a strong peak
in entanglement density on either side of the interface at $|z|\sim 10a$.
This was attributed to the higher density of chain ends and pancake like conformation produced by the
pre-welding interface.
This interpretation is bolstered by the lack of a similar peak in our healing simulations where the chain
conformations are not perturbed by reflections at an interface.

As the interdiffusion time $t$ increases, chains form new entanglements across the 
interface (Fig. \ref{fig:PPA}) and the dip in $\rho_{TC}$ fills in. At $t=7M\tau$, $\rho_{TC}$ is statistically indistinguishable from the bulk density all across the sample. Note that the entanglement density saturates more rapidly in welding samples, where
$\rho_{TC}$ reaches the bulk value near $2.5M\tau$ \cite{ge13}.
This more rapid evolution of entanglements contrasts with the slower evolution of $<d>$ in welding, and reflects the fact that diffusion by long chains is needed to produce entanglements.

Figure \ref{fig:TCs}(b) shows the rise of interfacial TCs at the interface during healing.
The areal density $N_{TC}^I/A$, which is shown in the inset for both the healing and welding samples, has been shown \cite{ge13} to directly correlate with the interfacial strength of welds.
As shown above for structural quantities characterizing the interfacial loops, $N_{TC}^I/A$ is slightly larger for the healing sample due to the faster diffusion of short chains.
Also shown in the inset are the areal densities of TCs with $n_{end}>50$ and $n_{end}>70$.
At the same $t$, $N_{TC}^I/A$ with $n_{end}>50$ is almost the same for healing and welding, while $N_{TC}^I/A$ with $n_{end}>70$ is lower for healing.
The results above are for samples of flexible chains.
Similar entanglement evolution across the interface is observed for $k_{\rm bend}>0$.
However, as the interdiffusion is slower for larger $k_{\rm bend}$, it takes longer for $\rho_{TC}$ to saturate at the corresponding bulk density.

\begin{figure}[!htbp]
\centerline{\includegraphics[width=0.4\textwidth]{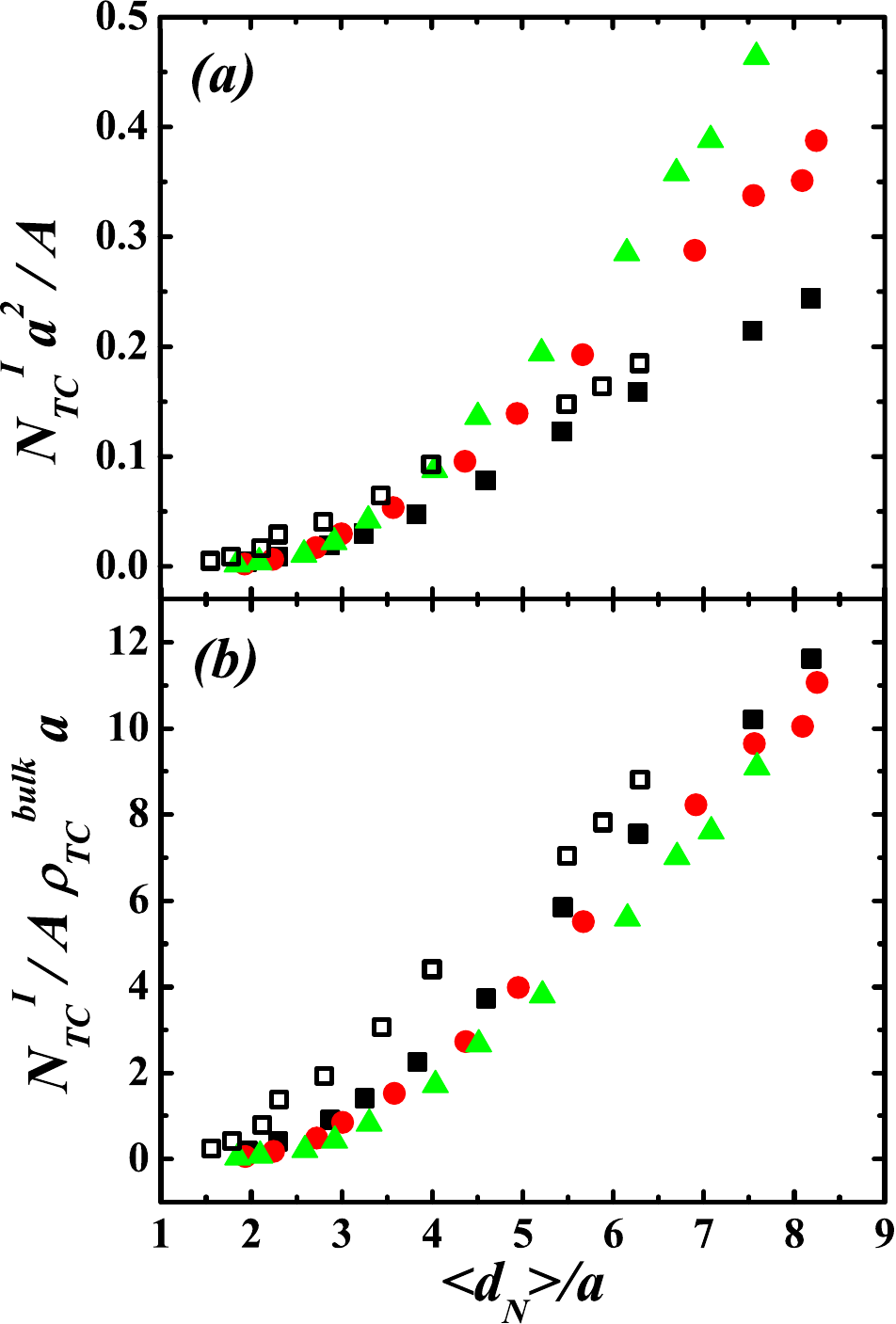}}
\caption{ (a) The areal density  of interfacial TCs and (b) the ratio of areal and bulk densities versus $\left< d_N \right>$, the average interpenetration depth of monomers weighted by the segment length $N$. Only interior TCs that are more than $30$ monomers from a chain end are included. Results are shown for the healing samples with $k_{\rm bend}/u_0=0$ (black squares), $0.75$ (red circles) and $1.5$ (green triangles), and the welding sample with $k_{\rm bend}=0$ (open black squares).
}
\label{fig:TC_depth}
\end{figure}

Previous simulations of thermal welding found an almost linear correlation between the areal density of interfacial TCs and the interpenetration depth above a threshold interpenetration depth \cite{ge13,ge13b}. For healing, this correlation is complicated by the polydispersity in chain length. To reduce the effects of short cut chains, we calculated the average interpenetration depth of monomers using the chain length $N$ of the segments they belong to as the weight. Figure \ref{fig:TC_depth}(a) plots the density of interfacial TCs 
with $n_{end}>30$ against the $N$ weighted average interpenetration depth $\left< d_N\right>$ of monomers. 
As found previously \cite{ge13,ge13b},
for welding with $k_{bend}=0$
the increase of $N_{TC}^I/A$ with $\left< d_N\right>$ is almost linear for all non-vanishing $N_{TC}^I/A$.
In contrast, for healing with $k_{bend}=0$, the increase is slow at small $\left< d_N\right>$, and becomes more rapid and linear only after $\left< d_N\right>$ rises above $\sim3.5a$. The reason is that short chain segments dominate the diffusion at small times but they contribute less significantly to the formation of TCs.
 
For healing samples, as $k_{bend}$ increases, the rise of $N_{TC}^I/A$ becomes more rapid. This is consistent with the results shown in Sec.~\ref{sec:method} that bulk samples of stiffer chains are more entangled. To separate the effects of chain stiffness and the interpenetration depth, we normalize $N_{TC}^I/A$ by the corresponding bulk density $\rho_{TC}^{bulk}$ of TCs. As shown in Fig.~\ref{fig:TC_depth}(b), the behavior of $N_{TC}^I/A\rho_{TC}^{bulk}$ versus $\left<d_N\right>$ is almost the same for different healing samples.
This result indicates that the different rates of increase in Fig.~\ref{fig:TC_depth}(a)
merely reflect the larger bulk entanglement density of stiffer chains.

\subsection{Interfacial Strength}
\label{sec:strength}
\subsubsection{\label{sec:bulk} Shear Response of Uncut Samples}

The ultimate goal of healing or welding is to recover the mechanical
strength of a bulk polymer.
This section begins by describing this bulk response, including the role
of stiffness and chain length.
These results are then
used to calibrate the response of healed and welded samples.

\begin{figure}[!htbp]
\begin{center}{
\includegraphics[width=0.45\textwidth]{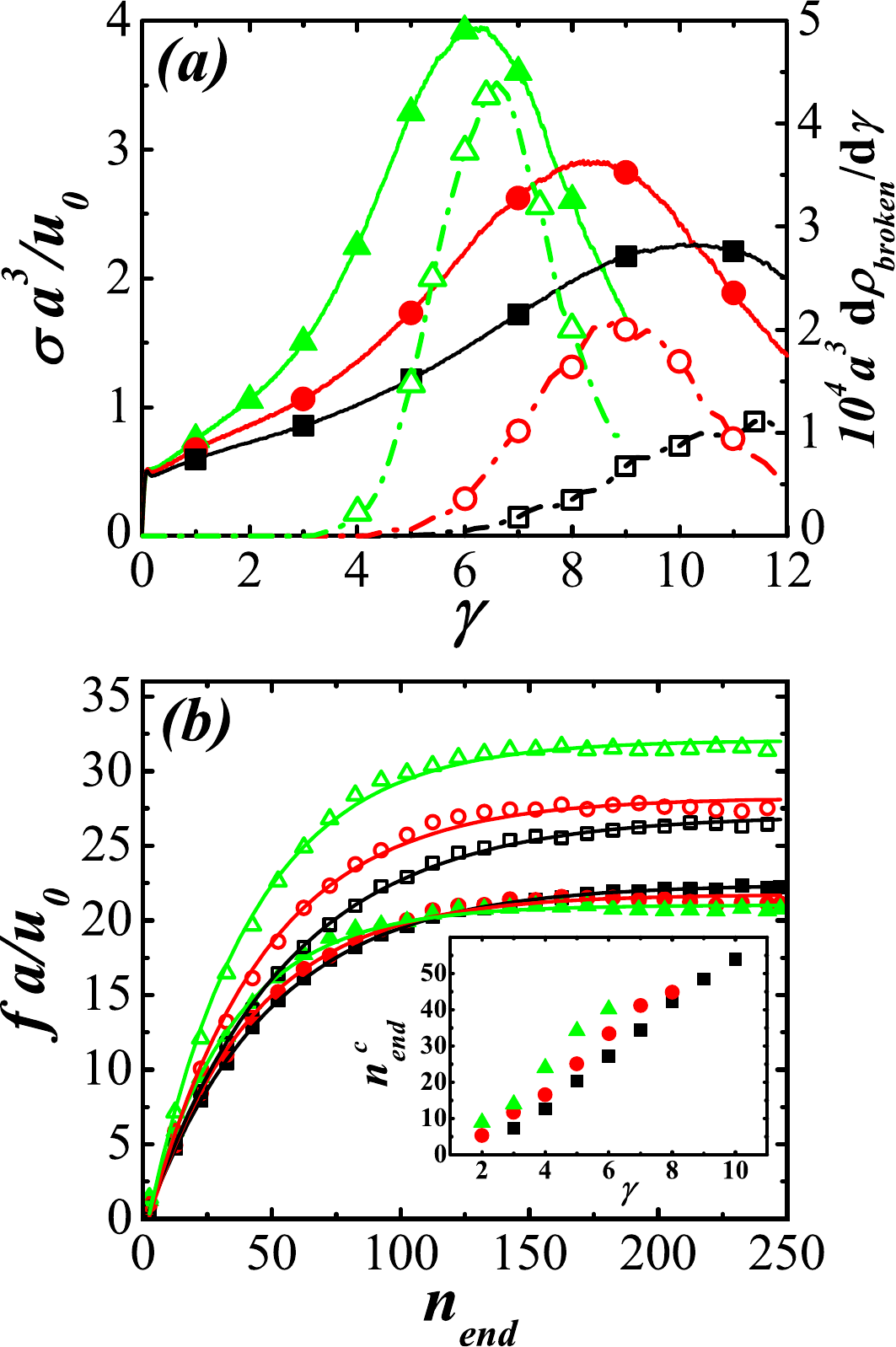}
}
\end{center}
\caption{(a) Average shear stress vs. shear strain curves (solid lines and symbols) for bulk samples with $k_{\rm bend}/u_0=0$ (black squares), $0.75$ (red circles) and $1.5$ (green triangles).
Dashed lines and open symbols show the rate of bond breaking $d \rho_{broken}/d\gamma$ where $\rho_{broken}$ is the density of broken bonds.
(b) Average bond tension $f$ along the chain as a function of $n_{end}$ for bulk samples with different $k_{\rm bend}$ at the shear strain $\gamma^c$ where significant bond breaking initiates (solid symbols) and $\gamma^{max}$ where the shear stress reaches its peak value (open symbols). For $k_{\rm bend}/u_0=0$, $0.75$ and $1.5$, $\gamma^c=9$, $7$ and $5$, and $\gamma^{max}=10$, $8$ and $6$ respectively. Solid lines are results from fitting the corresponding data points using $f=f_0(1-\exp(-n_{end}/n_{end}^c))$. The increase of $n_{end}^c$ with shear strain $\gamma$ is shown in the inset of (b) for different $k_{\rm bend}$.
}
\label{fig:stress_k}
\end{figure}

Typical stress-strain curves for samples with different chain stiffness are shown in Fig.~\ref{fig:stress_k}(a).
In all cases there is an initial elastic response followed by yield and strain softening.
This initial region is confined to $\gamma < 0.2$ and is difficult to see on the scale of the plot.
Further increases in $\gamma$ lead to strain hardening, a steady rise in the stress with strain.
Eventually the stress becomes high enough to produce chain scission.
The stress then rises less rapidly with strain, reaches a maximum, and 
finally drops as strain localizes on a failure plane.

Chain stiffness has little effect on the initial elastic response or
yield stress, but greatly enhances strain hardening \cite{hoy07,hoy08}.
The reason is that an affine shear displacement of monomers causes chains to stretch.
Stiffer, straighter chains
are less able to stretch without lengthening or breaking the strong covalent backbone bonds.
This has traditionally been modeled as a loss of entropy of the random walks
between entanglements \cite{haward68,arruda93b}, but recent work shows that the stress rises because the rate of plastic activity required to avoid bond breaking rises as chains straighten \cite{hoy07,hoy08,chen09b,ge10}.
Since stiffer chains cannot be straightened as much without stretching backbone bonds, the strain
$\gamma^c$ where significant chain scission occurs and the strain
$\gamma^{\rm max}$ where the stress peaks become smaller as $k_{\rm bend}$ increases.
Figure \ref{fig:stress_k}(a) also shows the rate of bond breaking.
We identify $\gamma^c =9$, $7$, and $5$ for $k_{\rm bend}=0$, $0.75$ and $1.5$
with the strain where the rate of acceleration in bond breaking
is highest.

While most of the energy associated with shearing the sample is dissipated as heat in plastic rearrangements, there is also a rise in the energy and the tension along backbone bonds.
Figure \ref{fig:stress_k}(b) shows the dependence of the average tension  on the chemical distance $n_{end}$ from the nearest chain end.
Only bonds in unbroken chains are included in the average and
results for $\gamma^c$ and $\gamma^{max}$ are shown.
In all cases, the average bond tension relaxes near chain ends, which are free to retract along their tube.
This relaxation can be fitted by $f=f_0 (1-\exp(-n_{end}/n_{end}^c))$, where $f_0$
corresponds to the plateau tension away from chain ends, while $n_{end}^c$ characterizes the distance over which $f$ relaxes \cite{rottler03}.
Both the plateau tension and the characteristic relaxation length $n_{end}^c$ increase with $\gamma$.

At $\gamma^{c}$, chains of different stiffness have nearly the same plateau bond tension, $\sim 22 u_0/a$.
This is much smaller than the bond breaking force $\sim 240 u_0/a$,
but the distribution of $f$ has an exponential tail like that seen
in craze formation \cite{rottler02b}.
The fraction of bonds that can break at any instant is of order
$\exp(-240/22) \approx 2 \times 10^{-5}$ and about 1 in $10^4$ has broken by $\gamma^c$ for all chain stiffnesses.
There is also a common value of $f_0 \sim 8 u_0/a$ at the start
of bond breaking, which is at strains of about 6, 5 and 3.5 for progressively stiffer chains.

The plateau bond tension at $\gamma^{max}$ increases with $k_{\rm bend}$ and the number of broken bonds also increases.
This suggests that as entanglement density increases with $k_{\rm bend}$, more bond breaking is needed to form a fracture plane across the whole sample.
We find that the total number of broken bonds per entanglement is
about a quarter at $\sigma^{max}$ and a little over unity at the point of final failure.

The increase of $n_{end}^c$ with $\gamma$ is shown explicitly in the inset of Fig.~\ref{fig:stress_k}(b).
The characteristic relaxation length $n_{end}^c$ at $\gamma^{c}$ decreases with
increasing $k_{\rm bend}$ ($48$, $41$ and $34$), but
the ratio of $n_{end}^c$ to $N_e$ increases,
$n_{end}^c/N_e\sim0.56$, $1.05$ and $1.31$. 
This indicates that entanglements alone do not determine $n_{end}^c$
and that the friction between chains required to retract them along their tube
is also important.
This conclusion is bolstered by considering the effect of chain length.

The above results are all for well-entangled chains of length $500$.
To estimate the potential effect of polydispersity near the cut interface,
we performed shear tests on monodispersed bulk samples of chains with different lengths.
Cubic samples with $\sim 3 \cdot 10^5$ beads and periodic boundary conditions ($L_x=L_y=L_z\sim70a$) were equilibrated as in Ref. \cite{auhl03}.
We tested that results for this reduced system size gave the same trends with $N$ as the larger systems used in healing and welding simulations.
There is a slight ($\sim 3$\%) decrease in maximum stress with system size
that may be related to the greater statistical likelihood of a weaker region in a larger system.

\begin{figure}[!htbp]
\centerline{\includegraphics[width=0.4\textwidth]{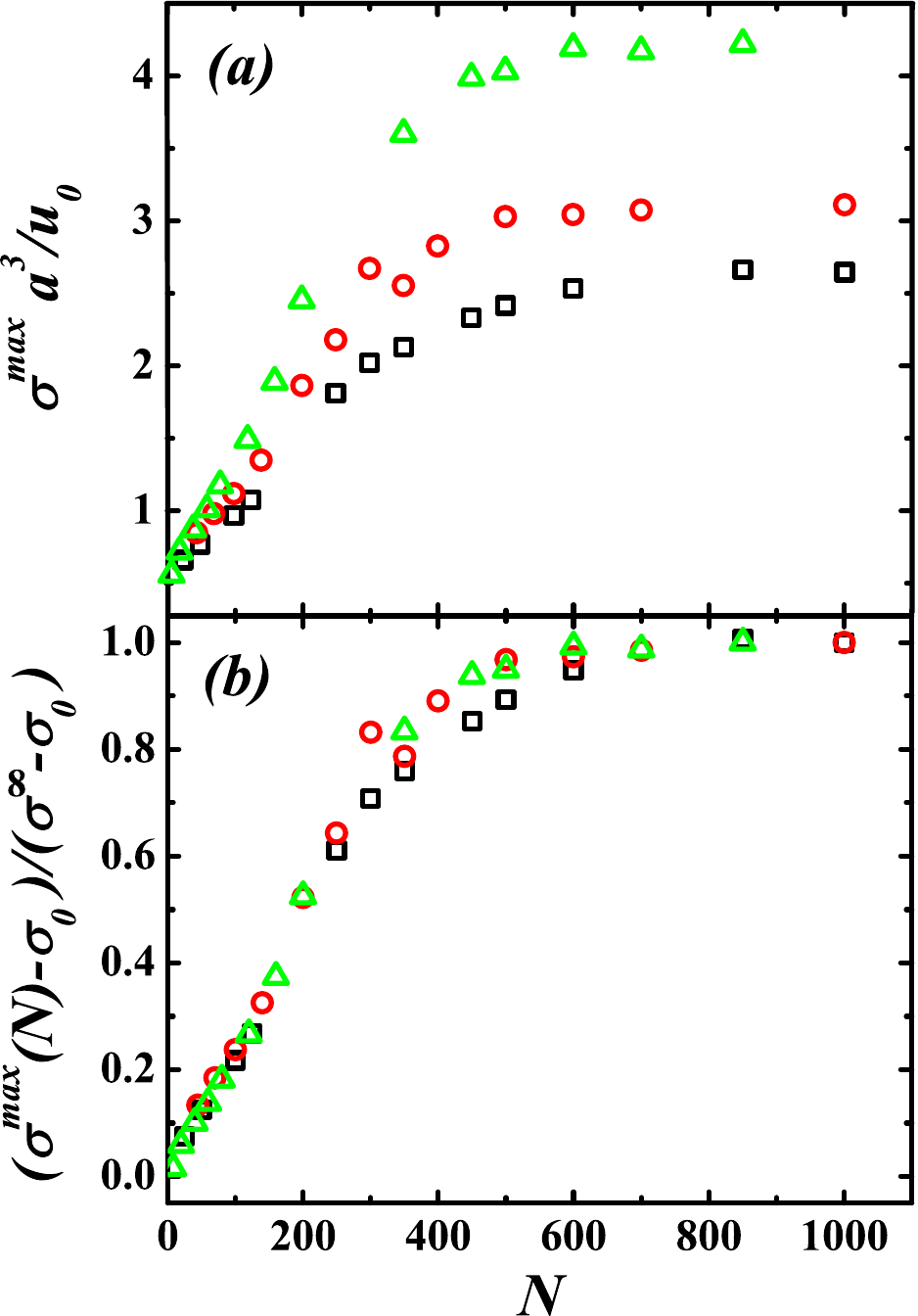}}
\caption{(a) The maximum shear stress $\sigma^{max}$ for monodispersed bulk samples with different chain lengths for $k_{\rm bend}/u_0=0$ (black squares), 0.75 (red circles) and 1.5 (green triangles).
(b) Rescaled plot showing the percentage of the increase with chain length as a function of $N$.
}
\label{fig:length}
\end{figure}

The chain length dependence of the bulk failure stress is shown in Fig.~\ref{fig:length}(a).
The shear strength $\sigma^{max}$ increases rapidly with $N$ and then saturates at a limiting long chain value which rises with increasing chain stiffness.
Even though $N_e$ decreases from about 85 to 26 with increasing $k_{\rm bend}$,
the asymptotic strength is reached at roughly the same chain length.
Figure \ref{fig:length}(b) shows a scaled plot of the percentage increase toward the bulk value.
About 90\% of the change in strength is complete by $N=400$ for the stiffest chains and by $N=500$ for the most flexible.
This combined with the plots of backbone tension in Fig. \ref{fig:stress_k}(b) suggest that interchain friction plays a critical role in determining when failure changes from pullout to scission.
Only when scission is dominant does the strength become independent of $N$.
Previous studies of tensile failure through crazing \cite{rottler03} had found
that the fracture energy of systems with different $k_{bend}$ saturated at similar values of $N/N_e$, indicating that chain friction was not important.
However the accepted values of $N_e$ at the time were
64 and 32 instead of 85 and 26 for $k_{bend}=0$ and $1.5$.
Reexamination of the data shows that the shift in the transition is smaller
than the change in $N_e$.

Interchain friction is difficult to measure or control in experiments,
so strength saturation is normally expressed in terms of $N/N_e$.
Using a 90\% criterion in Fig.~\ref{fig:length}, one obtains $N/N_e$ between 6 and 15.
Experimental studies of tensile fracture and interfacial strength for a range of polymers typically show saturation for a similar range of $N/N_e$ \cite{adolf85,mikos88,schnell98,wool95}. 

For the welding samples, all chains have $N=500$ and the strength is close to the long chain limit even for flexible chains.
In contrast, many of the chains near the interface of healed samples have
chains in the range from 150 to 400 where strength varies rapidly with length. As we will see below this leads to a lower shear strength.

In Sec.~\ref{sec:method} we noted that other forms of the covalent bonding potential had been used and the results were found to follow the same trends.
For any reasonable stiffness, the elastic response is dominated by the intermolecular interactions.
The main effect of the covalent potential comes from the breaking force $f_B$.
As $f_B$ decreases, the chain tension $f_0$ becomes large enough to break bonds
at smaller strains (Fig.~\ref{fig:stress_k}).
The ratio of the strength of long to short chains decreases, as does the increase in weld strength with time.
If $f_B$ is reduced by a factor of 3 to $80 u_0/a$, the difference between the failure stress of chains with $N_e$ and $N\rightarrow \infty$ is only 50\%.
For $f_B=160 u_0/a$ the limiting large $N$ failure stress is only $\sim 15$\% below that for $f_B=240u_0$, and further increases have less effect.
The factor of 3 increase in strength with weld time found here and in Ref. \cite{ge13} is similar to that found in experiments.

\subsubsection{\label{sec:timestrength} Time Dependent Strength of Healed and Welded Interfaces}

\begin{figure}[!htb]
\includegraphics[width=0.45\textwidth]{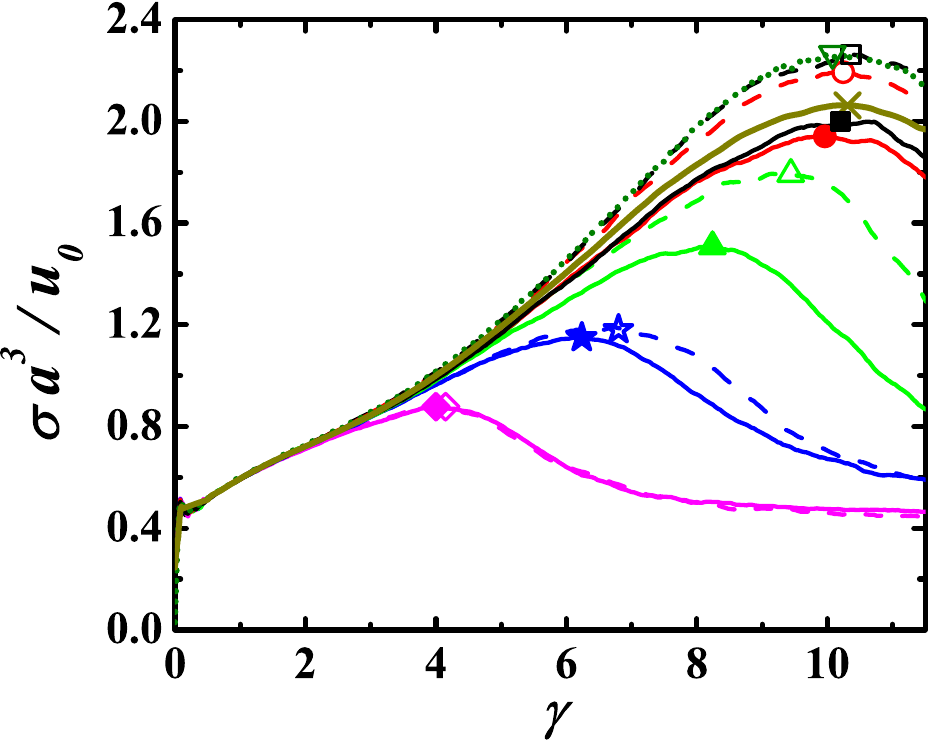}
\caption{Stress-strain curves from shear tests on different states for healing (solid lines and symbols) and welding (dashed lines and open symbols) of fully flexible chains ($k_{\rm bend}=0$). Symbols indicate the peak stress for each curve. Results are shown for interdiffusion time $t=0.01M\tau$ (magenta diamonds), $0.1M\tau$ (blue stars), $0.5M\tau$ (green triangles), $2M\tau$ (red circles), $5M\tau$ (black squares) and $16M\tau$ (dark yellow cross, only for healing). Also shown is the corresponding average stress-strain curve for the bulk (olive dotted line and inverted triangle).
}
\label{fig:strength}
\end{figure}

Figure \ref{fig:strength} contrasts the behavior of healed and welded samples
of flexible chains ($k_{\rm bend}=0$).
In all cases the initial elastic response, yield and strain hardening
follows the bulk response.
The youngest interfaces ($g=0.01M\tau$) begin to deviate from the bulk response near $\gamma =3$ and curves for healing and welding are nearly the same.
The peak stress of $\sim 0.8 u_0/a^3$ is comparable to values for bulk
simulations with unentangled chains (Fig. \ref{fig:length}),
which is consistent with the small number of interfacial entanglements and short loop lengths described in the previous sections.
As time increases, the results follow the bulk curve to larger strains
and the welding results for $t > 2.5M\tau$ are indistinguishable from the bulk response.
The curves for healing rise more slowly with time than welding and continue
to fall below the bulk response at $16M\tau$.

Figure~\ref{fig:snapshot} shows the evolution in the interfacial
structure before and after failure with increasing healing time.
Monomers that are initially below and above the cutting plane are
colored yellow and blue, respectively.
At short times the distribution of the two types of monomer sharpens with
increasing strain.
In Fig.~\ref{fig:snapshot}(d), many of the chains that had diffused
only part way across the interface have been pulled back to their initial side.
A few small chains that had diffused completely across the interface remain on the other side and no chain scission has occurred.
The final shear stress corresponds to friction between two polymer blocks with
sharp interfaces.

As the healing time increases, longer segments of chains span between the two sides. Many break rather than being pulled out, and the height distribution
of monomers from the two sides changes very little during failure.
The strong shear alignment of individual chains that leads to high stresses and scission is clearly visible.
Similar changes are observed in welding, but because there is
no polydispersity, interdiffusion is
slower and chain pullout produces sharper interfaces \cite{ge13}.

\begin{figure}[!htbp]
\centerline{\includegraphics[width=0.4\textwidth]{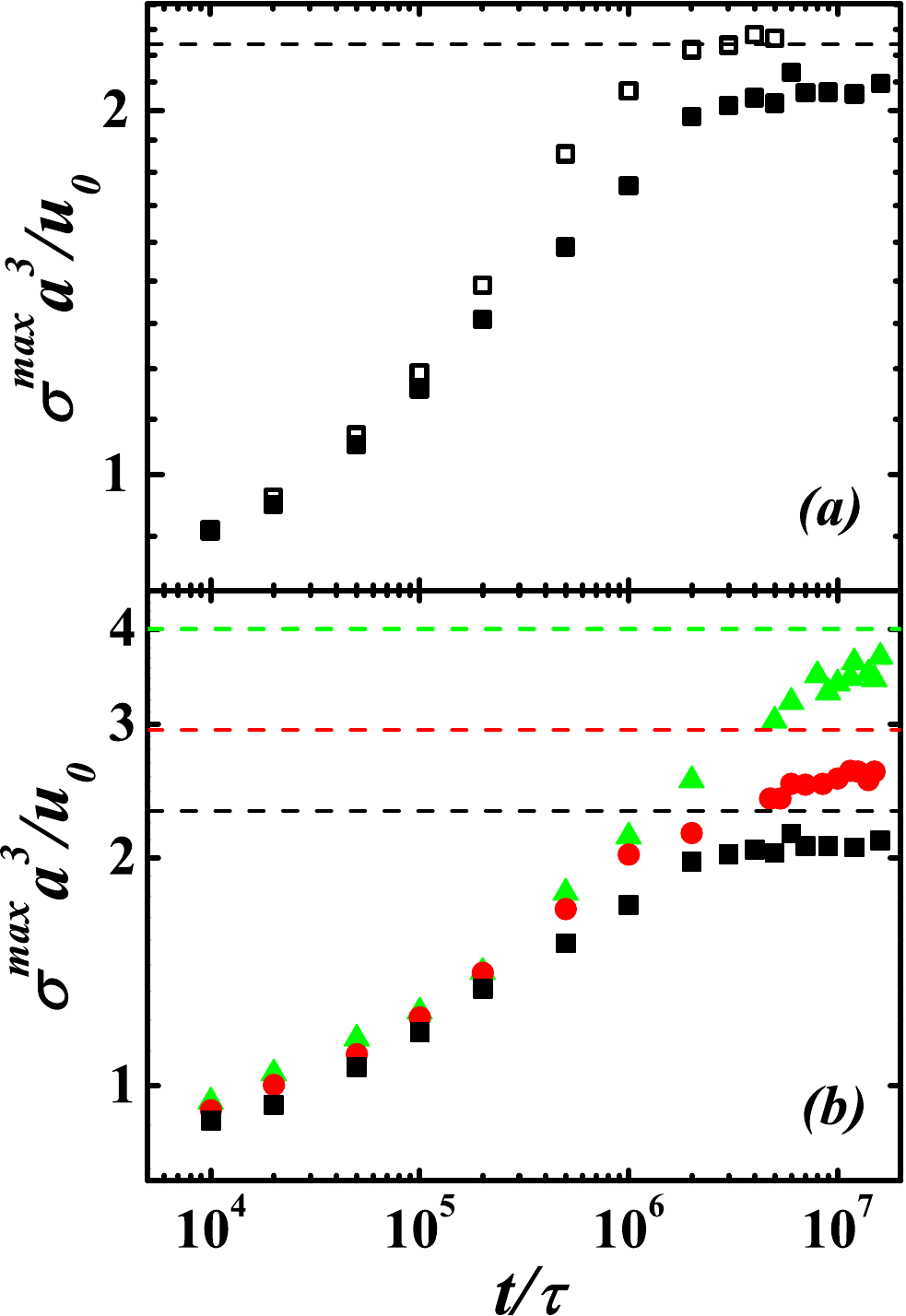}}
\caption{(a) The maximum shear stress $\sigma^{max}$ is shown versus $t$ for both healing (solid squares) and welding (open squares) samples of fully flexible chains.
(b) Time dependence of $\sigma^{max}$ for healing samples
with $k_{\rm bend}/u_0=0$ (black squares), $0.75$ (red circles) and $1.5$ (green triangles).
Dotted lines indicate the corresponding bulk strengths.
}
\label{fig:strength_time}
\end{figure}

As in previous studies \cite{ge13} and experiments \cite{kline88}, we use the maximum shear stress $\sigma^{max}$ before failure to characterize the interfacial strength.
Figure \ref{fig:strength_time}(a) shows $\sigma^{max}$ versus time for the healing and welding samples of Fig.~\ref{fig:strength}.
Both curves rise smoothly and appear to saturate for $t > 2M\tau$.
The welding results saturate at the bulk strength, while healing results are lower at all times and never reach the bulk strength.

Figure \ref{fig:strength_time}(b) shows healing curves for chains with different stiffness.
Results for different $k_{\rm bend}$ are almost identical for
$t \leq 2 \cdot 10^5$.
At this stage of interdiffusion, the amount of entanglements across the interface is low in all systems and the interface fails through chain pullout.
The force required to pull short chain segments from the opposite side is determined by interchain friction, which does not depend significantly on $k_{\rm bend}$.
This is also why the initial yield stress is similar for all $k_{\rm bend}$.
Note that the strain at which systems fail does depend on stiffness.
Because they strain harden more rapidly, stiffer systems reach the stress
required for chain pullout at a lower strain.

At later times the interfacial strength rises slightly more rapidly for stiffer
chains.
As the number of interfacial entanglements increases, chain pullout is arrested
and the interface begins to fail through chain scission.
The interface is progressively coupled to more of the surrounding bulk material whose strength grows with stiffness.
Ultimately, the interfacial strength saturates slightly below the bulk
strength for all three $k_{\rm bend}$.
The saturation time is longer for stiffer chains because of the slower diffusion: about $2M\tau$, $5M\tau$ and $10M\tau$ for $k_{\rm bend}/u_0=0$, $0.75$ and $1.5$, respectively. 
To check that the reduced strength is associated with the interface,
we performed additional simulations in wider systems ($H=110a$). 
At failure, strain was localized in the interfacial region
and the density of broken bonds was higher there.
Welded samples that have recovered bulk strength show uniform distributions of strain and broken bonds.

The reduced strength of healing samples can be directly related to the residual concentration of cut chains at the interface (Fig. \ref{fig:density}).
For the flexible chains, the plateau in $\sigma^{max}$ is comparable to the bulk
strength of chains of length 3 to $4N_e$.
A substantial fraction of interfacial chains have lengths in this range at $2M\tau$.
There is also a residual density of very short, unentangled chains that would be smaller in an infinite system where they could continue to diffuse.
To evaluate their effect we removed all chains with length less than 100 from the $7M\tau$ sample just before quenching. 
We found that this did not affect $\sigma^{max}$, implying that chains of length
100 to 500 are responsible for the lower interfacial stress.
Since diffusion of these chains away from the interface is slowed by entanglements, the ultimate recovery of strength to bulk values would take much longer times than are accessible in simulations.

These results suggest that recovery of the bulk strength at the interface in a healing sample is a two-step process.
This arises from the separation of two time scales: the time $\tau_1$ for the entanglement density to approach its bulk distribution and the time $\tau_2$ for all cut chains to diffuse away from the interface.
As shown in Fig. \ref{fig:TCs}, $\tau_1$ is much smaller than the disentanglement time $\tau_d$, while $\tau_2$ is expected to be multiple times $\tau_d$.
The strength begins to plateau near $\tau_1$ but full strength will not
be recovered until $\tau_2$. 
Similar two-step healing processes based on separation of diffusion time scales have been observed in the healing behavior of $\gamma$-irradiated poly(styrene-co-acrylonitrile) (SAN) \cite{nguyen81} and blends of polystyrene (PS) with poly(-2,6 dimethyl 1,4 phenylene oxide) (PPO) \cite{kausch87}.

\subsubsection{Relation Between Interfacial Strength and Structure}

\begin{figure}[!htb]
\centerline{\includegraphics[width=0.4\textwidth]{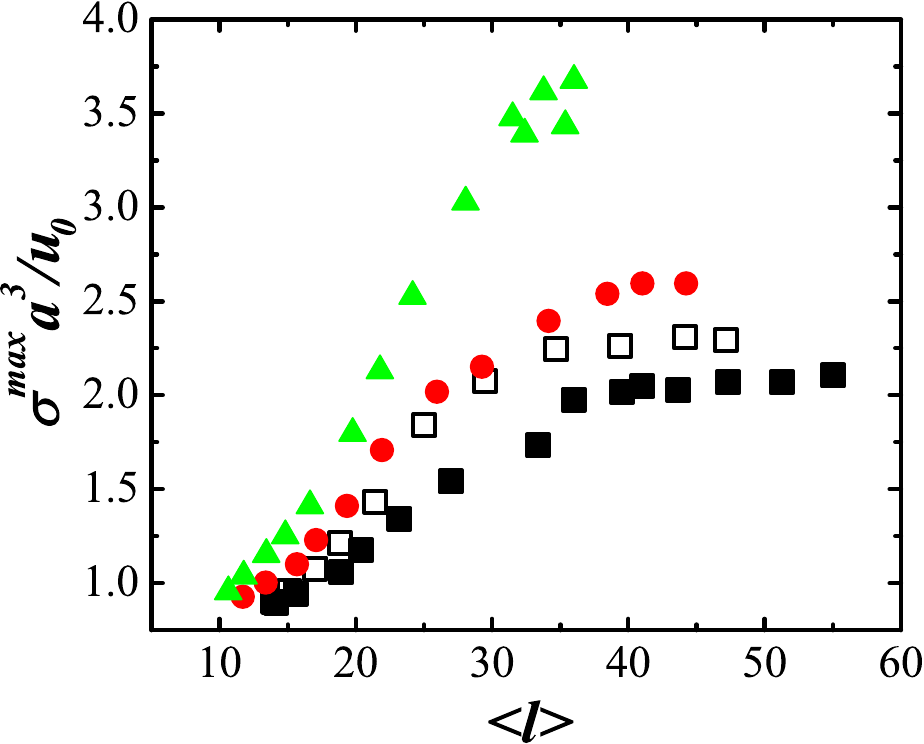}}
\caption{The maximum shear stress $\sigma^{max}$ versus the average contour length $\left< l \right>$ of interfacial loops for welding of flexible chains (open squares) and healing with $k_{\rm bend}/u_0=0$ (filled squares), 0.75 (circles) and 1.5 (triangles).
}
\label{fig:stress_loop}
\end{figure}

As noted above, models of healing and welding have typically related shear strength to equilibrium structural properties.
Given the trends in Sec. \ref{sec:structure}, it is clear that no
simple structural measure will collapse the different results in Fig.~\ref{fig:strength_time} for welding and healing.
The mean length of loops $\left< l \right>$ is nearly the same for welding and healing samples (Fig.~\ref{fig:loops}(a)) and the mean depth of loops $\left< h \right>$ is nearly the same for all systems (Fig. \ref{fig:loops}(b)).
Figure~\ref{fig:stress_loop} shows that plotting $\sigma^{max}$ 
against $\left< l \right>$ rather than time increases the separation between results for different stiffness, although they seem to saturate at similar values of $\left< l \right> \sim 35$.
The strength imparted by loops clearly depends on their stiffness and on
the length of the chains that anchor them into the opposing surface.

\begin{figure}[!htbp]
\centerline{\includegraphics[width=0.4\textwidth]{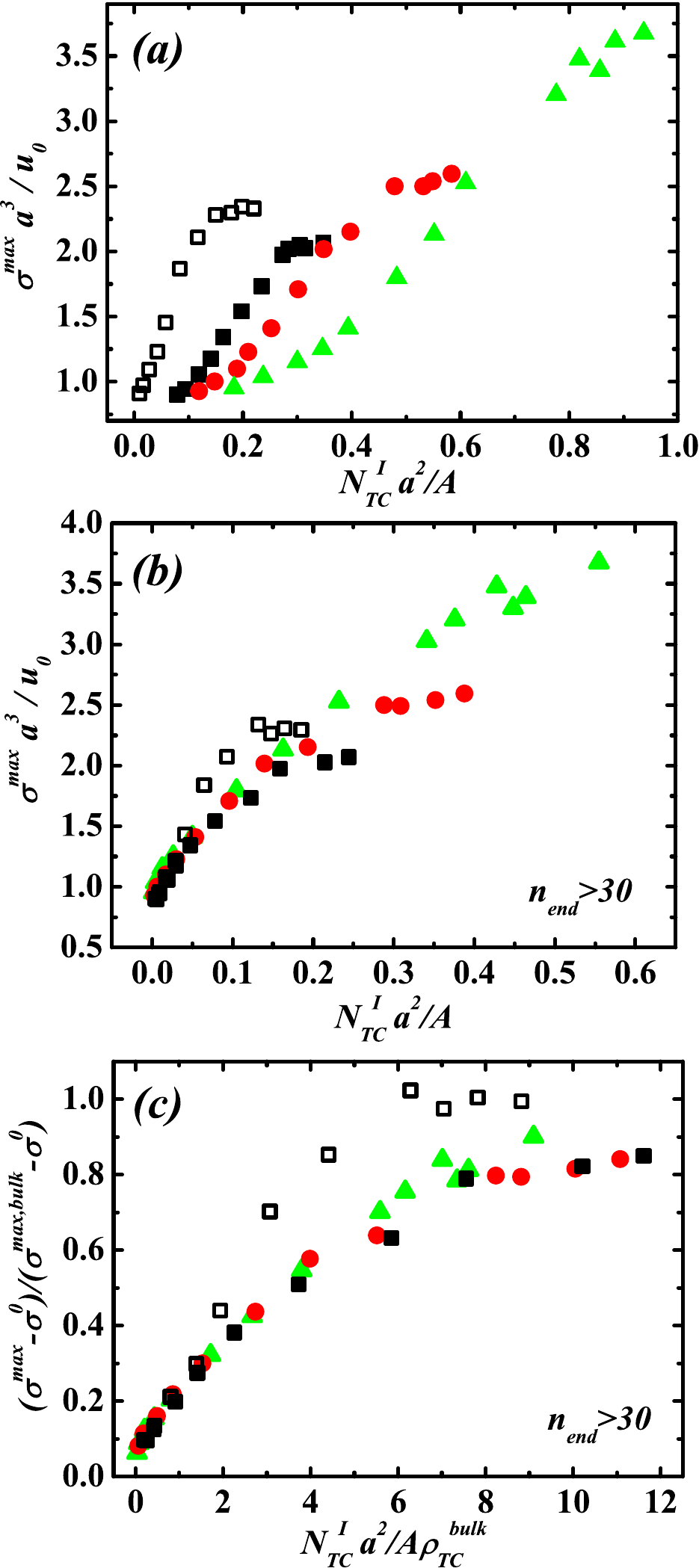}}
\caption{(a) Maximum shear stress $\sigma^{max}$ versus the areal density of interfacial TCs for healing (filled symbols) and welding (open symbols) with $k_{\rm bend}/u_0=0$ (squares), $k_{\rm bend}/u_0=0.75$ (circles) and
$k_{\rm bend}$ (triangles).
(b) Same plot but excluding entanglements less than 30 monomers from a chain end.
(c) Same data showing fraction of increase toward bulk strength against
areal density of interfacial entanglements normalized by bulk density.
}
\label{fig:stress_ITC}
\end{figure}

Our recent simulations \cite{ge13} of welding showed that the recovery of bulk shear strength coincided with the approach of the entanglement density at the interface to the bulk value.
Figure \ref{fig:stress_ITC}(a) shows the rise in strength with entanglement density for welding and healing samples.
As with other measures, the connection between strength and entanglement density is less clear for healing because of polydispersity.
Many of the entanglements involve short chains and can not impart as much strength to the interface.

We have tried a number of ways of correcting the entanglement density to reflect the greater effectiveness of longer chains.
From Fig.~\ref{fig:stress_k} it is clear that stress is able to relax near chain ends, and thus entanglements near chain ends are not effective in imparting strength.
Excluding entanglements less than $n_{0}$ from a chain end ($n_{end}<n_0$) also eliminates all entanglements involving chains of length less than $2 n_{0}$.
Fig.~\ref{fig:stress_ITC}(b) shows results for $n_0 =30$ which is roughly
the distance from the chain end where the stress reaches $f_0/2$ at $\gamma_c$
for all chain stiffnesses (Fig.~\ref{fig:stress_k}).
Results for all healing systems follow a common curve until they saturate at
a strength that grows with stiffness.
Welding results follow the same curve at low strengths, but then increase more
rapidly and saturate at a higher level.
This appears to reflect the fact that chain lengths of order $500$ are needed to reach the full bulk strength.
Some shorter chains are present near the interface for all healing systems.

Figure \ref{fig:TCs}(b) strongly supports the notion that the strength imparted by entanglements is proportional to the areal density of interfacial entanglements that are far enough from chain ends that they do not pull out.
As expected from Fig.~\ref{fig:stress_k}, taking $n_0$ much smaller than 30 includes too many entanglements that can pull out and does not collapse results for different stiffness. As in Fig.~\ref{fig:TCs}(a), $\rho_{TC}^I$ does not go to zero at the earliest times because of motion by short chains during the quench to the glassy state.
Increasing $n_0$ to 50 excludes too many entanglements and there
is a significant increase in strength before there are any interfacial entanglements, particularly for large $k_{\rm bend}$.

The bulk density of entanglements increases with chain stiffness and it is interesting
to see if that correlates with the increasing strength of stiffer chains.
Fig.~\ref{fig:TCs}(c) shows the fractional increase from the short chain limit, $\sigma_0 \sim 0.8 u_0/a^3$, to the bulk strength normalized by the bulk entanglement density.
Note that this collapses all of the healing results onto a single curve and the welding results follow the same curve at early times.
This collapse is consistent with the idea that recovery towards the bulk entanglement density is related to recovery to bulk strength.

The ratio of interfacial areal density to bulk density has units of length, indicating that there is a common depth that entanglements must span to recover bulk strength. Indeed, as shown in Fig.~\ref{fig:TC_depth}(b), for healing samples with different $k_{bend}$, the same $N_{TC}^I/A\rho_{TC}^{bulk}$ corresponds to almost the same $\left<d_N\right>$. A common interdiffusion depth for strength recovery is consistent with strength saturation at a constant loop length in Fig.~\ref{fig:stress_loop} and a common length for stress relaxation with all $k_{\rm bend}$. It also explains why it takes longer for the interfacial strength to saturate for larger $k_{\rm bend}$. Since the interfacial dynamics slows down with increasing $k_{\rm bend}$, the time it takes to reach the same depth increases.

Experiments \cite{schnell98,schnell99,brown01,benkoski01} show that there is a direct correlation between the width $a_I$ of the interface and the interfacial fracture energy $G$. However, the increase of $G/G^{max}$ with $a_I/d_T$, where $G^{max}$ is the saturation value of $G$ and $d_T$ the tube diameter, is material specific. In particular, at $G_{max}$, $a_I$ varies between 1 and 2 times $d_T$. Benkoski {\it et al.} \cite{benkoski02} argued that changes of interchain friction can fully account for this non-universal behavior. 

In our simulations, the ratio of the common interdiffusion depth at strength recovery to the corresponding tube diameter varies with $k_{bend}$. As in experiments \cite{schnell98,schnell99,brown01,benkoski01}, this result suggests that entanglements alone do not fully determine the strength recovery at the interface. The role of entanglements is coupled with that of interchain friction, which has been demonstrated in Fig.~\ref{fig:stress_k}(b).   

\section{Summary}
\label{sec:sum}

Extensive simulations of the thermal healing of precut polymer samples were performed. The precut introduces a symmetric polymer-polymer interface with polydispersity in chain length. This polydispersity complicates the interdiffusion. The unentangled cut chains quickly diffuse away from the interface, and after a relatively short time are almost uniformly distributed across the finite simulation box. Cut chains longer than the entanglement length diffuse less rapidly, and are still predominantly distributed around the interface at large diffusion times.
These chains slow the development of bulk strength even though the mass uptake and increase in interpenetration depth are faster than for thermal
welding of monodispersed polymers. 

We examine the evolution of the interfacial structure during healing and welding.
The increase of the mean length $\left<l\right>$ and depth $\left<d\right>$ of interfacial loops with interdiffusion time is almost the same for healing and welding samples.
The mean length rises slightly more slowly for healing of stiffer chains because they diffuse more slowly.
However, since the conformation of stiffer chains is more extended,
$\left<d\right>$ is nearly independent of stiffness.
The number of interfacial loops is nearly independent of time for both healing
and welding, in sharp contrast to the $t^{1/4}$ increase predicted by previous theoretical work \cite{zhang89}.

Primitive path analysis provided direct information about entanglements across the interface. The number of interfacial entanglements increases faster during healing than welding. However, a larger fraction of interfacial entanglements in healing samples are near chain ends and the number of entanglements in the center of chains rises more slowly than for welding.
As a result, the density of entanglements near the interface takes longer
to recover to the bulk density.
Experiments often measure the interdiffusion depth as a proxy for entanglement density \cite{schnell98,schnell99,brown01}.
The primitive path analysis shows
the number of interfacial entanglements rises more rapidly with interdiffusion depth for healing of stiffer chains, but the results collapse when normalized by the bulk entanglement density.
The interdiffusion depth needed to produce a significant number of interfacial entanglements is roughly twice as big for healing as welding.

A simple shear test is applied to measure the shear strength of different samples. The strength is quantified using the maximum shear stress $\sigma^{max}$ before failure. We first study the effects of chain stiffness and length on the bulk shear response. As chain stiffness $k_{bend}$ increases, strain hardening is enhanced. As a result, the stress level at any given strain increases. Moreover, the strain $\gamma^c$ where significant chain scission occurs and the strain $\gamma^{max}$ that corresponds to $\sigma^{max}$ both decrease with $k_{bend}$. At $\gamma^c$, the average bond tension $f_0$ near the center of chains and the relaxation distance $n_{end}^c$ from chain ends are independent of $k_{bend}$.
However, the value of $f_0$ at $\gamma^{max}$ increases with $k_{bend}$.
The resulting increase in the number of broken bonds is correlated to the increase in entanglement density.
In all cases, one bond is broken for every four entanglements at $\gamma^{max}$ and by the final failure about one bond is broken for each entanglement.
The mean bond tension remains an order of magnitude smaller than the breaking force and scission is associated with bonds in the tail of an exponential
distribution of local tensions.

As chain length increases, the bulk $\sigma^{max}$ first increases and then saturates at the long chain limit. Chain stiffness increases the limiting value. However, the limit is approached at almost the same chain length for different $k_{bend}$. This suggests that interchain friction, which is independent of stiffness, is critical in determining the chain length for which chain pullout dominates over chain scission. 
More evidence comes from trends in the length $n_{end}^c$ over which bond tension relaxes through pullout at chain ends.
At a given $f_0$, $n_{end}^c$ decreases only slightly (~30\%) with stiffness for healing samples while the ratio $n_{end}^c/N_e$ more than doubles.

The time dependence of shear strength at healing and welding interfaces is contrasted.
For healing, the bulk mechanical response is not fully recovered on the time scales of our simulations. In contrast to welding, $\sigma^{max}$ saturates below the bulk value and 
the rise of $\sigma^{max}$ before saturation is slower.
We attribute the retarded strength development to the presence of cut chains with $N>N_e$ at the interface, which are less entangled than the uncut chains. The ultimate recovery of bulk strength relies on the diffusion of these entangled cut chains away from the interface, which would take a much longer time and therefore is inaccessible to our simulations. 

Chain stiffness affects the development of strength in healing samples. At early times failure is through chain pullout. The value of $\sigma^{max}$ is almost independent of $k_{bend}$ because pullout stresses are mainly determined by interchain friction.
Later on, $\sigma^{max}$ rises more rapidly and reaches a higher saturation value for larger $k_{\rm bend}$, consistent with the higher stress curves and $\sigma^{max}$ in the bulk.
Moreover, the saturation occurs at a larger time for larger $k_{\rm bend}$ due to slower interdiffusion. 

The relation between interfacial strength development and interfacial structure evolution is examined.
Neither the interdiffusion depth, the mean length $\left<l\right>$ of interfacial loops nor $N_{TC}^{I}/A$ give neat predictions for the relative strength development in healing and welding. Nevertheless, for healing with different $k_{bend}$, a good correlation is found between the fractional increase of $\sigma^{max}$ from the short chain limit to the bulk strength and the areal density of interfacial entanglements divided by the bulk density $N_{TC}^I/A\rho_{TC}^{bulk}$. This indicates that the average interpenetration depth at strength saturation is almost the same for different $k_{\rm bend}$, consistent with the universal correlation between $N_{TC}^I/A\rho_{TC}^{bulk}$ and $\left<d_N\right>$. The ratio of the common interpenetration depth to the corresponding tube diameter varies for different $k_{bend}$. Once again, this suggests that entanglements and interchain friction are coupled to strengthen the interface.  

\section*{\label{sec:ack} Acknowledgement}
We thank E. J. Kramer and M. Rubinstein for useful discussions. This work was supported by the National Science Foundation under grants DMR-1006805, CMMI-0923018, and OCI-0963185.
MOR acknowledges support from the Simons Foundation. D.P. and G.S.G. acknowledge support from Department of Energy Award No. DE-FG02-12ER46843. This research used resources at the National Energy Research Scientific Computing Center (NERSC), which is supported by the Office of Science of the United States Department of Energy under Contract No. DE-AC02-05CH11231. Research was carried out in part, at the Center for Integrated Nanotechnologies, a U.S. Department of Energy, Office of Basic Energy Sciences user facility. Sandia National Laboratories is a multi-program laboratory managed and operated by Sandia Corporation, a wholly owned subsidiary of Lockheed Martin Corporation, for the U.S. Department of Energy's National Nuclear Security Administration under contract DE-AC04-94AL85000. 

\bibliography{self-healing14}

\begin{thebibliography}{70}
\expandafter\ifx\csname natexlab\endcsname\relax\def\natexlab#1{#1}\fi
\expandafter\ifx\csname bibnamefont\endcsname\relax
  \def\bibnamefont#1{#1}\fi
\expandafter\ifx\csname bibfnamefont\endcsname\relax
  \def\bibfnamefont#1{#1}\fi
\expandafter\ifx\csname citenamefont\endcsname\relax
  \def\citenamefont#1{#1}\fi
\expandafter\ifx\csname url\endcsname\relax
  \def\url#1{\texttt{#1}}\fi
\expandafter\ifx\csname urlprefix\endcsname\relax\def\urlprefix{URL }\fi
\providecommand{\bibinfo}[2]{#2}
\providecommand{\eprint}[2][]{\url{#2}}

\bibitem[{\citenamefont{Wool}(1995)}]{wool95}
\bibinfo{author}{\bibfnamefont{R.~P.} \bibnamefont{Wool}},
  \emph{\bibinfo{title}{Polymer Interfaces: Structure and Strength}}
  (\bibinfo{publisher}{Hanser}, \bibinfo{address}{Munich},
  \bibinfo{year}{1995}).

\bibitem[{\citenamefont{Haward and Young}(1995)}]{haward97}
\bibinfo{author}{\bibfnamefont{R.~N.} \bibnamefont{Haward}} \bibnamefont{and}
  \bibinfo{author}{\bibfnamefont{R.~J.} \bibnamefont{Young}},
  \emph{\bibinfo{title}{The Physics of Glassy Polymers}}
  (\bibinfo{publisher}{Chapman \& Hall}, \bibinfo{address}{London},
  \bibinfo{year}{1995}).

\bibitem[{\citenamefont{Jones and Richards}(1999)}]{jone99}
\bibinfo{author}{\bibfnamefont{R.~A.~L.} \bibnamefont{Jones}} \bibnamefont{and}
  \bibinfo{author}{\bibfnamefont{R.~W.} \bibnamefont{Richards}},
  \emph{\bibinfo{title}{Polymers at Surfaces and Interfaces}}
  (\bibinfo{publisher}{Cambridge University Press}, \bibinfo{address}{New
  York}, \bibinfo{year}{1999}).

\bibitem[{\citenamefont{Jud et~al.}(1981)\citenamefont{Jud, Kausch, and
  Williams}}]{jud81}
\bibinfo{author}{\bibfnamefont{K.}~\bibnamefont{Jud}},
  \bibinfo{author}{\bibfnamefont{H.~H.} \bibnamefont{Kausch}},
  \bibnamefont{and} \bibinfo{author}{\bibfnamefont{J.~G.}
  \bibnamefont{Williams}}, \bibinfo{journal}{J. Mater. Sci.}
  \textbf{\bibinfo{volume}{16}}, \bibinfo{pages}{204} (\bibinfo{year}{1981}).

\bibitem[{\citenamefont{Wool and O'Connor}(1981)}]{wool81}
\bibinfo{author}{\bibfnamefont{R.~P.} \bibnamefont{Wool}} \bibnamefont{and}
  \bibinfo{author}{\bibfnamefont{K.~M.} \bibnamefont{O'Connor}},
  \bibinfo{journal}{J. Appl. Phys.} \textbf{\bibinfo{volume}{52}},
  \bibinfo{pages}{5953} (\bibinfo{year}{1981}).

\bibitem[{\citenamefont{Prager and Tirrell}(1981)}]{prager81}
\bibinfo{author}{\bibfnamefont{S.}~\bibnamefont{Prager}} \bibnamefont{and}
  \bibinfo{author}{\bibfnamefont{M.}~\bibnamefont{Tirrell}},
  \bibinfo{journal}{J. Chem. Phys.} \textbf{\bibinfo{volume}{75}},
  \bibinfo{pages}{5194} (\bibinfo{year}{1981}).

\bibitem[{\citenamefont{Prager et~al.}(1983)\citenamefont{Prager, Adolf, and
  Tirrell}}]{prager83}
\bibinfo{author}{\bibfnamefont{S.}~\bibnamefont{Prager}},
  \bibinfo{author}{\bibfnamefont{D.}~\bibnamefont{Adolf}}, \bibnamefont{and}
  \bibinfo{author}{\bibfnamefont{M.}~\bibnamefont{Tirrell}},
  \bibinfo{journal}{J. Chem. Phys.} \textbf{\bibinfo{volume}{78}},
  \bibinfo{pages}{7015} (\bibinfo{year}{1983}).

\bibitem[{\citenamefont{Kim and Wool}(1983)}]{kim83}
\bibinfo{author}{\bibfnamefont{Y.~H.} \bibnamefont{Kim}} \bibnamefont{and}
  \bibinfo{author}{\bibfnamefont{R.~P.} \bibnamefont{Wool}},
  \bibinfo{journal}{Macromolecules} \textbf{\bibinfo{volume}{16}},
  \bibinfo{pages}{1115} (\bibinfo{year}{1983}).

\bibitem[{\citenamefont{Adolf et~al.}(1985)\citenamefont{Adolf, Tirrell, and
  Prager}}]{adolf85}
\bibinfo{author}{\bibfnamefont{D.}~\bibnamefont{Adolf}},
  \bibinfo{author}{\bibfnamefont{M.}~\bibnamefont{Tirrell}}, \bibnamefont{and}
  \bibinfo{author}{\bibfnamefont{S.}~\bibnamefont{Prager}},
  \bibinfo{journal}{J. Polym. Sci., Polym. Phys. Ed.}
  \textbf{\bibinfo{volume}{23}}, \bibinfo{pages}{413} (\bibinfo{year}{1985}).

\bibitem[{\citenamefont{Mikos and Peppas}(1989)}]{mikos89}
\bibinfo{author}{\bibfnamefont{A.~G.} \bibnamefont{Mikos}} \bibnamefont{and}
  \bibinfo{author}{\bibfnamefont{N.~A.} \bibnamefont{Peppas}},
  \bibinfo{journal}{Polymer} \textbf{\bibinfo{volume}{30}}, \bibinfo{pages}{84}
  (\bibinfo{year}{1989}).

\bibitem[{\citenamefont{Russell et~al.}(1993)\citenamefont{Russell, Deline,
  Dozier, Felcher, Agrawal, Wool, and Mays}}]{russell93}
\bibinfo{author}{\bibfnamefont{T.~R.} \bibnamefont{Russell}},
  \bibinfo{author}{\bibfnamefont{V.~R.} \bibnamefont{Deline}},
  \bibinfo{author}{\bibfnamefont{W.~D.} \bibnamefont{Dozier}},
  \bibinfo{author}{\bibfnamefont{G.~P.} \bibnamefont{Felcher}},
  \bibinfo{author}{\bibfnamefont{G.}~\bibnamefont{Agrawal}},
  \bibinfo{author}{\bibfnamefont{R.~P.} \bibnamefont{Wool}}, \bibnamefont{and}
  \bibinfo{author}{\bibfnamefont{J.~W.} \bibnamefont{Mays}},
  \bibinfo{journal}{Nature} \textbf{\bibinfo{volume}{365}},
  \bibinfo{pages}{235} (\bibinfo{year}{1993}).

\bibitem[{\citenamefont{Wool}(2008)}]{wool08}
\bibinfo{author}{\bibfnamefont{R.~P.} \bibnamefont{Wool}},
  \bibinfo{journal}{Soft Matter} \textbf{\bibinfo{volume}{4}},
  \bibinfo{pages}{400} (\bibinfo{year}{2008}).

\bibitem[{\citenamefont{Blaiszik et~al.}(2010)\citenamefont{Blaiszik, Kramer,
  Olugebefola, Moore, Sottos, and White}}]{blaiszik10}
\bibinfo{author}{\bibfnamefont{B.~J.} \bibnamefont{Blaiszik}},
  \bibinfo{author}{\bibfnamefont{S.~L.~B.} \bibnamefont{Kramer}},
  \bibinfo{author}{\bibfnamefont{S.~C.} \bibnamefont{Olugebefola}},
  \bibinfo{author}{\bibfnamefont{J.~S.} \bibnamefont{Moore}},
  \bibinfo{author}{\bibfnamefont{N.~R.} \bibnamefont{Sottos}},
  \bibnamefont{and} \bibinfo{author}{\bibfnamefont{S.~R.} \bibnamefont{White}},
  \bibinfo{journal}{Annu. Rev. Mater. Res.} \textbf{\bibinfo{volume}{40}},
  \bibinfo{pages}{179} (\bibinfo{year}{2010}).

\bibitem[{\citenamefont{de~Gennes}(1971)}]{degennes71}
\bibinfo{author}{\bibfnamefont{P.~G.} \bibnamefont{de~Gennes}},
  \bibinfo{journal}{J. Chem. Phys.} \textbf{\bibinfo{volume}{55}},
  \bibinfo{pages}{572} (\bibinfo{year}{1971}).

\bibitem[{\citenamefont{Doi and Edwards}(1988)}]{doi88}
\bibinfo{author}{\bibfnamefont{M.}~\bibnamefont{Doi}} \bibnamefont{and}
  \bibinfo{author}{\bibfnamefont{S.~F.} \bibnamefont{Edwards}},
  \emph{\bibinfo{title}{The Theory of Polymer Dynamics}}
  (\bibinfo{publisher}{Oxford University Press}, \bibinfo{address}{Oxford},
  \bibinfo{year}{1988}).

\bibitem[{\citenamefont{Schnell et~al.}(1999)\citenamefont{Schnell, Stamm, and
  Creton}}]{schnell99}
\bibinfo{author}{\bibfnamefont{R.}~\bibnamefont{Schnell}},
  \bibinfo{author}{\bibfnamefont{M.}~\bibnamefont{Stamm}}, \bibnamefont{and}
  \bibinfo{author}{\bibfnamefont{C.}~\bibnamefont{Creton}},
  \bibinfo{journal}{Macromolecules} \textbf{\bibinfo{volume}{32}},
  \bibinfo{pages}{3420} (\bibinfo{year}{1999}).

\bibitem[{\citenamefont{Benkoski et~al.}(2002)\citenamefont{Benkoski,
  Fredrickson, and Kramer}}]{benkoski02}
\bibinfo{author}{\bibfnamefont{J.~J.} \bibnamefont{Benkoski}},
  \bibinfo{author}{\bibfnamefont{G.~H.} \bibnamefont{Fredrickson}},
  \bibnamefont{and} \bibinfo{author}{\bibfnamefont{E.~J.}
  \bibnamefont{Kramer}}, \bibinfo{journal}{J. Polym. Sci., Part B: Polym.
  Phys.} \textbf{\bibinfo{volume}{40}}, \bibinfo{pages}{2377}
  (\bibinfo{year}{2002}).

\bibitem[{\citenamefont{Kunz and Stamm}(1996)}]{kunz96}
\bibinfo{author}{\bibfnamefont{K.}~\bibnamefont{Kunz}} \bibnamefont{and}
  \bibinfo{author}{\bibfnamefont{M.}~\bibnamefont{Stamm}},
  \bibinfo{journal}{Macromolecules} \textbf{\bibinfo{volume}{29}},
  \bibinfo{pages}{2548} (\bibinfo{year}{1996}).

\bibitem[{\citenamefont{Schnell et~al.}(1998)\citenamefont{Schnell, Stamm, and
  Creton}}]{schnell98}
\bibinfo{author}{\bibfnamefont{R.}~\bibnamefont{Schnell}},
  \bibinfo{author}{\bibfnamefont{M.}~\bibnamefont{Stamm}}, \bibnamefont{and}
  \bibinfo{author}{\bibfnamefont{C.}~\bibnamefont{Creton}},
  \bibinfo{journal}{Macromolecules} \textbf{\bibinfo{volume}{31}},
  \bibinfo{pages}{2284} (\bibinfo{year}{1998}).

\bibitem[{\citenamefont{Brown}(2001)}]{brown01}
\bibinfo{author}{\bibfnamefont{H.~R.} \bibnamefont{Brown}},
  \bibinfo{journal}{Macromolecules} \textbf{\bibinfo{volume}{34}},
  \bibinfo{pages}{3720} (\bibinfo{year}{2001}).

\bibitem[{\citenamefont{McGraw et~al.}(2013)\citenamefont{McGraw, Fowler,
  Ferrari, and Dalnoki-Veress}}]{mcgraw13}
\bibinfo{author}{\bibfnamefont{J.~D.} \bibnamefont{McGraw}},
  \bibinfo{author}{\bibfnamefont{P.~D.} \bibnamefont{Fowler}},
  \bibinfo{author}{\bibfnamefont{M.~L.} \bibnamefont{Ferrari}},
  \bibnamefont{and}
  \bibinfo{author}{\bibfnamefont{K.}~\bibnamefont{Dalnoki-Veress}},
  \bibinfo{journal}{Eur. Phys. J. E} \textbf{\bibinfo{volume}{36}},
  \bibinfo{pages}{7} (\bibinfo{year}{2013}).

\bibitem[{\citenamefont{Kline and Wool}(1988)}]{kline88}
\bibinfo{author}{\bibfnamefont{D.~B.} \bibnamefont{Kline}} \bibnamefont{and}
  \bibinfo{author}{\bibfnamefont{R.~P.} \bibnamefont{Wool}},
  \bibinfo{journal}{Polym. Eng. Sci.} \textbf{\bibinfo{volume}{28}},
  \bibinfo{pages}{52} (\bibinfo{year}{1988}).

\bibitem[{\citenamefont{Parsons et~al.}(1998)\citenamefont{Parsons, Ernst,
  Smyser, Hiltner, and Baer}}]{parsons98}
\bibinfo{author}{\bibfnamefont{M.}~\bibnamefont{Parsons}},
  \bibinfo{author}{\bibfnamefont{A.}~\bibnamefont{Ernst}},
  \bibinfo{author}{\bibfnamefont{G.}~\bibnamefont{Smyser}},
  \bibinfo{author}{\bibfnamefont{A.}~\bibnamefont{Hiltner}}, \bibnamefont{and}
  \bibinfo{author}{\bibfnamefont{E.}~\bibnamefont{Baer}}, \bibinfo{journal}{J.
  Adhesion} \textbf{\bibinfo{volume}{66}}, \bibinfo{pages}{135}
  (\bibinfo{year}{1998}).

\bibitem[{\citenamefont{Akabori et~al.}(2006)\citenamefont{Akabori, Baba,
  Koguchi, Tanaka, and Nagamura}}]{akabori06}
\bibinfo{author}{\bibfnamefont{K.}~\bibnamefont{Akabori}},
  \bibinfo{author}{\bibfnamefont{D.}~\bibnamefont{Baba}},
  \bibinfo{author}{\bibfnamefont{K.}~\bibnamefont{Koguchi}},
  \bibinfo{author}{\bibfnamefont{K.}~\bibnamefont{Tanaka}}, \bibnamefont{and}
  \bibinfo{author}{\bibfnamefont{T.}~\bibnamefont{Nagamura}},
  \bibinfo{journal}{J. Polym. Sci., Part B: Polym. Phys.}
  \textbf{\bibinfo{volume}{44}}, \bibinfo{pages}{3589} (\bibinfo{year}{2006}).

\bibitem[{\citenamefont{Boiko}(2012)}]{boiko12}
\bibinfo{author}{\bibfnamefont{Y.~M.} \bibnamefont{Boiko}},
  \bibinfo{journal}{Macromol. Symp.} \textbf{\bibinfo{volume}{316}},
  \bibinfo{pages}{71} (\bibinfo{year}{2012}).

\bibitem[{\citenamefont{Helfand and Tagami}(1971)}]{helfand71}
\bibinfo{author}{\bibfnamefont{E.}~\bibnamefont{Helfand}} \bibnamefont{and}
  \bibinfo{author}{\bibfnamefont{Y.}~\bibnamefont{Tagami}},
  \bibinfo{journal}{J. Polym. Sci., Part B: Polym. Phys.}
  \textbf{\bibinfo{volume}{9}}, \bibinfo{pages}{741} (\bibinfo{year}{1971}).

\bibitem[{\citenamefont{Helfand and Tagami}(1972)}]{helfand72}
\bibinfo{author}{\bibfnamefont{E.}~\bibnamefont{Helfand}} \bibnamefont{and}
  \bibinfo{author}{\bibfnamefont{Y.}~\bibnamefont{Tagami}},
  \bibinfo{journal}{J. Chem. Phys.} \textbf{\bibinfo{volume}{56}},
  \bibinfo{pages}{3592} (\bibinfo{year}{1972}).

\bibitem[{\citenamefont{Deutsch and Binder}(1991)}]{deutsch91}
\bibinfo{author}{\bibfnamefont{H.~P.} \bibnamefont{Deutsch}} \bibnamefont{and}
  \bibinfo{author}{\bibfnamefont{K.}~\bibnamefont{Binder}},
  \bibinfo{journal}{J. Chem. Phys.} \textbf{\bibinfo{volume}{94}},
  \bibinfo{pages}{2294} (\bibinfo{year}{1991}).

\bibitem[{\citenamefont{Haire and Windle}(2001)}]{haire01}
\bibinfo{author}{\bibfnamefont{K.~R.} \bibnamefont{Haire}} \bibnamefont{and}
  \bibinfo{author}{\bibfnamefont{A.~H.} \bibnamefont{Windle}},
  \bibinfo{journal}{Comput. Theor. Polym. Sci.} \textbf{\bibinfo{volume}{11}},
  \bibinfo{pages}{227} (\bibinfo{year}{2001}).

\bibitem[{\citenamefont{Anderson et~al.}(2004)\citenamefont{Anderson, Wescott,
  Carver, and Windle}}]{anderson04}
\bibinfo{author}{\bibfnamefont{K.~L.} \bibnamefont{Anderson}},
  \bibinfo{author}{\bibfnamefont{J.~T.} \bibnamefont{Wescott}},
  \bibinfo{author}{\bibfnamefont{T.~J.} \bibnamefont{Carver}},
  \bibnamefont{and} \bibinfo{author}{\bibfnamefont{A.~H.}
  \bibnamefont{Windle}}, \bibinfo{journal}{Materials Science \& Engineering A}
  \textbf{\bibinfo{volume}{365}}, \bibinfo{pages}{14} (\bibinfo{year}{2004}).

\bibitem[{\citenamefont{Pierce et~al.}(2011)\citenamefont{Pierce, Perahia, and
  Grest}}]{pierce11}
\bibinfo{author}{\bibfnamefont{F.}~\bibnamefont{Pierce}},
  \bibinfo{author}{\bibfnamefont{D.}~\bibnamefont{Perahia}}, \bibnamefont{and}
  \bibinfo{author}{\bibfnamefont{G.~S.} \bibnamefont{Grest}},
  \bibinfo{journal}{EPL} \textbf{\bibinfo{volume}{95}}, \bibinfo{pages}{46001}
  (\bibinfo{year}{2011}).

\bibitem[{\citenamefont{Ge et~al.}(2013{\natexlab{a}})\citenamefont{Ge, Pierce,
  Perahia, Grest, and Robbins}}]{ge13}
\bibinfo{author}{\bibfnamefont{T.}~\bibnamefont{Ge}},
  \bibinfo{author}{\bibfnamefont{F.}~\bibnamefont{Pierce}},
  \bibinfo{author}{\bibfnamefont{D.}~\bibnamefont{Perahia}},
  \bibinfo{author}{\bibfnamefont{G.~S.} \bibnamefont{Grest}}, \bibnamefont{and}
  \bibinfo{author}{\bibfnamefont{M.~O.} \bibnamefont{Robbins}},
  \bibinfo{journal}{Phys. Rev. Lett.} \textbf{\bibinfo{volume}{110}},
  \bibinfo{pages}{98301} (\bibinfo{year}{2013}{\natexlab{a}}).

\bibitem[{\citenamefont{Ge et~al.}(2013{\natexlab{b}})\citenamefont{Ge, Grest,
  and Robbins}}]{ge13b}
\bibinfo{author}{\bibfnamefont{T.}~\bibnamefont{Ge}},
  \bibinfo{author}{\bibfnamefont{G.~S.} \bibnamefont{Grest}}, \bibnamefont{and}
  \bibinfo{author}{\bibfnamefont{M.~O.} \bibnamefont{Robbins}},
  \bibinfo{journal}{ACS Macro Lett.} \textbf{\bibinfo{volume}{2}},
  \bibinfo{pages}{882} (\bibinfo{year}{2013}{\natexlab{b}}).

\bibitem[{\citenamefont{Everaers et~al.}(2004)\citenamefont{Everaers,
  Sukumaran, Grest, Svaneborg, Sivasubramanian, and Kremer}}]{everaers04}
\bibinfo{author}{\bibfnamefont{R.}~\bibnamefont{Everaers}},
  \bibinfo{author}{\bibfnamefont{S.~K.} \bibnamefont{Sukumaran}},
  \bibinfo{author}{\bibfnamefont{G.~S.} \bibnamefont{Grest}},
  \bibinfo{author}{\bibfnamefont{C.}~\bibnamefont{Svaneborg}},
  \bibinfo{author}{\bibfnamefont{A.}~\bibnamefont{Sivasubramanian}},
  \bibnamefont{and} \bibinfo{author}{\bibfnamefont{K.}~\bibnamefont{Kremer}},
  \bibinfo{journal}{Science} \textbf{\bibinfo{volume}{303}},
  \bibinfo{pages}{823} (\bibinfo{year}{2004}).

\bibitem[{\citenamefont{Kr\"oger}(2005)}]{kroger05}
\bibinfo{author}{\bibfnamefont{M.}~\bibnamefont{Kr\"oger}},
  \bibinfo{journal}{Comput. Phys. Commun.} \textbf{\bibinfo{volume}{168}},
  \bibinfo{pages}{209} (\bibinfo{year}{2005}).

\bibitem[{\citenamefont{Tzoumanekas and Theodorou}(2006)}]{tzoumanekas06}
\bibinfo{author}{\bibfnamefont{C.}~\bibnamefont{Tzoumanekas}} \bibnamefont{and}
  \bibinfo{author}{\bibfnamefont{D.~N.} \bibnamefont{Theodorou}},
  \bibinfo{journal}{Macromolecules} \textbf{\bibinfo{volume}{39}},
  \bibinfo{pages}{4592} (\bibinfo{year}{2006}).

\bibitem[{\citenamefont{Kremer and Grest}(1990)}]{kremer90}
\bibinfo{author}{\bibfnamefont{K.}~\bibnamefont{Kremer}} \bibnamefont{and}
  \bibinfo{author}{\bibfnamefont{G.~S.} \bibnamefont{Grest}},
  \bibinfo{journal}{J. Chem. Phys.} \textbf{\bibinfo{volume}{92}},
  \bibinfo{pages}{5057} (\bibinfo{year}{1990}).

\bibitem[{\citenamefont{Rottler et~al.}(2002)\citenamefont{Rottler, Barsky, and
  Robbins}}]{rottler02a}
\bibinfo{author}{\bibfnamefont{J.}~\bibnamefont{Rottler}},
  \bibinfo{author}{\bibfnamefont{S.}~\bibnamefont{Barsky}}, \bibnamefont{and}
  \bibinfo{author}{\bibfnamefont{M.~O.} \bibnamefont{Robbins}},
  \bibinfo{journal}{Phys. Rev. Lett.} \textbf{\bibinfo{volume}{89}},
  \bibinfo{pages}{148304} (\bibinfo{year}{2002}).

\bibitem[{\citenamefont{Sides et~al.}(2001)\citenamefont{Sides, Grest, and
  Stevens}}]{sides01}
\bibinfo{author}{\bibfnamefont{S.~W.} \bibnamefont{Sides}},
  \bibinfo{author}{\bibfnamefont{G.~S.} \bibnamefont{Grest}}, \bibnamefont{and}
  \bibinfo{author}{\bibfnamefont{M.~J.} \bibnamefont{Stevens}},
  \bibinfo{journal}{Phys. Rev. E} \textbf{\bibinfo{volume}{64}},
  \bibinfo{pages}{050802} (\bibinfo{year}{2001}).

\bibitem[{\citenamefont{Stevens}(2001{\natexlab{a}})}]{stevens01b}
\bibinfo{author}{\bibfnamefont{M.~J.} \bibnamefont{Stevens}},
  \bibinfo{journal}{Macromolecules} \textbf{\bibinfo{volume}{34}},
  \bibinfo{pages}{1411} (\bibinfo{year}{2001}{\natexlab{a}}).

\bibitem[{\citenamefont{Stevens}(2001{\natexlab{b}})}]{stevens01}
\bibinfo{author}{\bibfnamefont{M.~J.} \bibnamefont{Stevens}},
  \bibinfo{journal}{Macromolecules} \textbf{\bibinfo{volume}{34}},
  \bibinfo{pages}{2710} (\bibinfo{year}{2001}{\natexlab{b}}).

\bibitem[{\citenamefont{Odell and Keller}(1986)}]{odell86}
\bibinfo{author}{\bibfnamefont{J.~A.} \bibnamefont{Odell}} \bibnamefont{and}
  \bibinfo{author}{\bibfnamefont{A.}~\bibnamefont{Keller}},
  \bibinfo{journal}{J. Polym. Sci., Part B: Polym. Phys.}
  \textbf{\bibinfo{volume}{24}}, \bibinfo{pages}{1889} (\bibinfo{year}{1986}).

\bibitem[{\citenamefont{Creton et~al.}(1992)\citenamefont{Creton, Kramer, Hui,
  and Brown}}]{creton92}
\bibinfo{author}{\bibfnamefont{C.}~\bibnamefont{Creton}},
  \bibinfo{author}{\bibfnamefont{E.~J.} \bibnamefont{Kramer}},
  \bibinfo{author}{\bibfnamefont{C.~Y.} \bibnamefont{Hui}}, \bibnamefont{and}
  \bibinfo{author}{\bibfnamefont{H.~R.} \bibnamefont{Brown}},
  \bibinfo{journal}{Macromolecules} \textbf{\bibinfo{volume}{25}},
  \bibinfo{pages}{3075} (\bibinfo{year}{1992}).

\bibitem[{foo()}]{foot0}
\bibinfo{note}{T. O'Connor and J. Andzelm private communication}.

\bibitem[{\citenamefont{P\"utz et~al.}(2000)\citenamefont{P\"utz, Kremer, and
  Grest}}]{puetz00}
\bibinfo{author}{\bibfnamefont{M.}~\bibnamefont{P\"utz}},
  \bibinfo{author}{\bibfnamefont{K.}~\bibnamefont{Kremer}}, \bibnamefont{and}
  \bibinfo{author}{\bibfnamefont{G.~S.} \bibnamefont{Grest}},
  \bibinfo{journal}{Europhys. Lett.} \textbf{\bibinfo{volume}{49}},
  \bibinfo{pages}{735} (\bibinfo{year}{2000}).

\bibitem[{\citenamefont{Hoy et~al.}(2009)\citenamefont{Hoy, Foteinopoulou, and
  Kr\"oger}}]{hoy09}
\bibinfo{author}{\bibfnamefont{R.~S.} \bibnamefont{Hoy}},
  \bibinfo{author}{\bibfnamefont{K.}~\bibnamefont{Foteinopoulou}},
  \bibnamefont{and} \bibinfo{author}{\bibfnamefont{M.}~\bibnamefont{Kr\"oger}},
  \bibinfo{journal}{Phys. Rev. E} \textbf{\bibinfo{volume}{80}},
  \bibinfo{pages}{031803} (\bibinfo{year}{2009}).

\bibitem[{\citenamefont{Hou et~al.}(2010)\citenamefont{Hou, Svaneborg,
  Everaers, and Grest}}]{hou10}
\bibinfo{author}{\bibfnamefont{J.-X.} \bibnamefont{Hou}},
  \bibinfo{author}{\bibfnamefont{C.}~\bibnamefont{Svaneborg}},
  \bibinfo{author}{\bibfnamefont{R.}~\bibnamefont{Everaers}}, \bibnamefont{and}
  \bibinfo{author}{\bibfnamefont{G.~S.} \bibnamefont{Grest}},
  \bibinfo{journal}{Phys. Rev. Lett.} \textbf{\bibinfo{volume}{105}},
  \bibinfo{pages}{068301} (\bibinfo{year}{2010}).

\bibitem[{\citenamefont{Rottler and Robbins}(2003{\natexlab{a}})}]{rottler03}
\bibinfo{author}{\bibfnamefont{J.}~\bibnamefont{Rottler}} \bibnamefont{and}
  \bibinfo{author}{\bibfnamefont{M.~O.} \bibnamefont{Robbins}},
  \bibinfo{journal}{Phys. Rev. E} \textbf{\bibinfo{volume}{68}},
  \bibinfo{pages}{011801} (\bibinfo{year}{2003}{\natexlab{a}}).

\bibitem[{\citenamefont{Plimpton}(1995)}]{plimpton95}
\bibinfo{author}{\bibfnamefont{S.}~\bibnamefont{Plimpton}},
  \bibinfo{journal}{J. Comp. Phys.} \textbf{\bibinfo{volume}{117}},
  \bibinfo{pages}{1} (\bibinfo{year}{1995}).

\bibitem[{\citenamefont{Auhl et~al.}(2003)\citenamefont{Auhl, Everaers, Grest,
  Kremer, and Plimpton}}]{auhl03}
\bibinfo{author}{\bibfnamefont{R.}~\bibnamefont{Auhl}},
  \bibinfo{author}{\bibfnamefont{R.}~\bibnamefont{Everaers}},
  \bibinfo{author}{\bibfnamefont{G.~S.} \bibnamefont{Grest}},
  \bibinfo{author}{\bibfnamefont{K.}~\bibnamefont{Kremer}}, \bibnamefont{and}
  \bibinfo{author}{\bibfnamefont{S.~J.} \bibnamefont{Plimpton}},
  \bibinfo{journal}{J. Chem. Phys.} \textbf{\bibinfo{volume}{119}},
  \bibinfo{pages}{12718} (\bibinfo{year}{2003}).

\bibitem[{\citenamefont{Rottler and Robbins}(2002)}]{rottler02b}
\bibinfo{author}{\bibfnamefont{J.}~\bibnamefont{Rottler}} \bibnamefont{and}
  \bibinfo{author}{\bibfnamefont{M.~O.} \bibnamefont{Robbins}},
  \bibinfo{journal}{Phys. Rev. Lett.} \textbf{\bibinfo{volume}{89}},
  \bibinfo{pages}{195501} (\bibinfo{year}{2002}).

\bibitem[{\citenamefont{Hoy and Robbins}(2007)}]{hoy07}
\bibinfo{author}{\bibfnamefont{R.~S.} \bibnamefont{Hoy}} \bibnamefont{and}
  \bibinfo{author}{\bibfnamefont{M.~O.} \bibnamefont{Robbins}},
  \bibinfo{journal}{Phys. Rev. Lett.} \textbf{\bibinfo{volume}{99}},
  \bibinfo{pages}{117801} (\bibinfo{year}{2007}).

\bibitem[{\citenamefont{Hoy and Robbins}(2008)}]{hoy08}
\bibinfo{author}{\bibfnamefont{R.~S.} \bibnamefont{Hoy}} \bibnamefont{and}
  \bibinfo{author}{\bibfnamefont{M.~O.} \bibnamefont{Robbins}},
  \bibinfo{journal}{Phys. Rev. E} \textbf{\bibinfo{volume}{77}},
  \bibinfo{pages}{031801} (\bibinfo{year}{2008}).

\bibitem[{\citenamefont{Rottler and Robbins}(2005)}]{rottler05}
\bibinfo{author}{\bibfnamefont{J.}~\bibnamefont{Rottler}} \bibnamefont{and}
  \bibinfo{author}{\bibfnamefont{M.~O.} \bibnamefont{Robbins}},
  \bibinfo{journal}{Phys. Rev. Lett.} \textbf{\bibinfo{volume}{95}},
  \bibinfo{pages}{225504} (\bibinfo{year}{2005}).

\bibitem[{\citenamefont{Hasan and Boyce}(1993)}]{hasan93}
\bibinfo{author}{\bibfnamefont{O.~A.} \bibnamefont{Hasan}} \bibnamefont{and}
  \bibinfo{author}{\bibfnamefont{M.~C.} \bibnamefont{Boyce}},
  \bibinfo{journal}{Polymer} \textbf{\bibinfo{volume}{34}},
  \bibinfo{pages}{5085} (\bibinfo{year}{1993}).

\bibitem[{\citenamefont{Rottler and Robbins}(2003{\natexlab{b}})}]{rottler03c}
\bibinfo{author}{\bibfnamefont{J.}~\bibnamefont{Rottler}} \bibnamefont{and}
  \bibinfo{author}{\bibfnamefont{M.~O.} \bibnamefont{Robbins}},
  \bibinfo{journal}{Phys. Rev. E} \textbf{\bibinfo{volume}{68}},
  \bibinfo{pages}{011507} (\bibinfo{year}{2003}{\natexlab{b}}).

\bibitem[{\citenamefont{Hoy and Grest}(2007)}]{hoy07b}
\bibinfo{author}{\bibfnamefont{R.~S.} \bibnamefont{Hoy}} \bibnamefont{and}
  \bibinfo{author}{\bibfnamefont{G.~S.} \bibnamefont{Grest}},
  \bibinfo{journal}{Macromolecules} \textbf{\bibinfo{volume}{40}},
  \bibinfo{pages}{8389} (\bibinfo{year}{2007}).

\bibitem[{\citenamefont{Everaers}(2012)}]{everaers12}
\bibinfo{author}{\bibfnamefont{R.}~\bibnamefont{Everaers}},
  \bibinfo{journal}{Phys. Rev. E} \textbf{\bibinfo{volume}{86}},
  \bibinfo{pages}{022801} (\bibinfo{year}{2012}).

\bibitem[{\citenamefont{de~Gennes}(1989)}]{degennes89}
\bibinfo{author}{\bibfnamefont{P.~G.} \bibnamefont{de~Gennes}},
  \bibinfo{journal}{C. R. Acad. Sci. (Paris) Ser. II}
  \textbf{\bibinfo{volume}{308}}, \bibinfo{pages}{1401} (\bibinfo{year}{1989}).

\bibitem[{\citenamefont{Zhang and Wool}(1989)}]{zhang89}
\bibinfo{author}{\bibfnamefont{H.}~\bibnamefont{Zhang}} \bibnamefont{and}
  \bibinfo{author}{\bibfnamefont{R.~P.} \bibnamefont{Wool}},
  \bibinfo{journal}{Macromolecules} \textbf{\bibinfo{volume}{22}},
  \bibinfo{pages}{3018} (\bibinfo{year}{1989}).

\bibitem[{\citenamefont{Kim et~al.}(1996)\citenamefont{Kim, Lee, and
  Lee}}]{kim96}
\bibinfo{author}{\bibfnamefont{H.~J.} \bibnamefont{Kim}},
  \bibinfo{author}{\bibfnamefont{K.}~\bibnamefont{Lee}}, \bibnamefont{and}
  \bibinfo{author}{\bibfnamefont{H.~H.} \bibnamefont{Lee}},
  \bibinfo{journal}{Polymer} \textbf{\bibinfo{volume}{37}},
  \bibinfo{pages}{4593} (\bibinfo{year}{1996}).

\bibitem[{\citenamefont{Mikos and Peppas}(1988)}]{mikos88}
\bibinfo{author}{\bibfnamefont{A.~G.} \bibnamefont{Mikos}} \bibnamefont{and}
  \bibinfo{author}{\bibfnamefont{N.~A.} \bibnamefont{Peppas}},
  \bibinfo{journal}{J. Chem. Phys.} \textbf{\bibinfo{volume}{88}},
  \bibinfo{pages}{1337} (\bibinfo{year}{1988}).

\bibitem[{\citenamefont{Silvestri et~al.}(2003)\citenamefont{Silvestri, Brown,
  Carra, and Carra}}]{silvestri03}
\bibinfo{author}{\bibfnamefont{L.}~\bibnamefont{Silvestri}},
  \bibinfo{author}{\bibfnamefont{H.~R.} \bibnamefont{Brown}},
  \bibinfo{author}{\bibfnamefont{S.}~\bibnamefont{Carra}}, \bibnamefont{and}
  \bibinfo{author}{\bibfnamefont{S.}~\bibnamefont{Carra}}, \bibinfo{journal}{J.
  Chem. Phys.} \textbf{\bibinfo{volume}{119}}, \bibinfo{pages}{8140}
  (\bibinfo{year}{2003}).

\bibitem[{\citenamefont{Haward and Thackray}(1968)}]{haward68}
\bibinfo{author}{\bibfnamefont{R.~N.} \bibnamefont{Haward}} \bibnamefont{and}
  \bibinfo{author}{\bibfnamefont{G.}~\bibnamefont{Thackray}},
  \bibinfo{journal}{Proc. R. Soc. A} \textbf{\bibinfo{volume}{302}},
  \bibinfo{pages}{453} (\bibinfo{year}{1968}).

\bibitem[{\citenamefont{Arruda and Boyce}(1993)}]{arruda93b}
\bibinfo{author}{\bibfnamefont{E.~M.} \bibnamefont{Arruda}} \bibnamefont{and}
  \bibinfo{author}{\bibfnamefont{M.~C.} \bibnamefont{Boyce}},
  \bibinfo{journal}{Int. J. Plast.} \textbf{\bibinfo{volume}{9}},
  \bibinfo{pages}{697} (\bibinfo{year}{1993}).

\bibitem[{\citenamefont{Chen and Schweizer}(2009)}]{chen09b}
\bibinfo{author}{\bibfnamefont{K.}~\bibnamefont{Chen}} \bibnamefont{and}
  \bibinfo{author}{\bibfnamefont{K.~S.} \bibnamefont{Schweizer}},
  \bibinfo{journal}{Phys. Rev. Lett.} \textbf{\bibinfo{volume}{102}},
  \bibinfo{pages}{038301} (\bibinfo{year}{2009}).

\bibitem[{\citenamefont{Ge and Robbins}(2010)}]{ge10}
\bibinfo{author}{\bibfnamefont{T.}~\bibnamefont{Ge}} \bibnamefont{and}
  \bibinfo{author}{\bibfnamefont{M.~O.} \bibnamefont{Robbins}},
  \bibinfo{journal}{J. Polym. Sci., Part B: Polym. Phys.}
  \textbf{\bibinfo{volume}{48}}, \bibinfo{pages}{1473} (\bibinfo{year}{2010}).

\bibitem[{\citenamefont{Nguyen et~al.}(1981)\citenamefont{Nguyen, Kausch, Jud,
  and Dettenmaier}}]{nguyen81}
\bibinfo{author}{\bibfnamefont{T.~Q.} \bibnamefont{Nguyen}},
  \bibinfo{author}{\bibfnamefont{H.~H.} \bibnamefont{Kausch}},
  \bibinfo{author}{\bibfnamefont{K.}~\bibnamefont{Jud}}, \bibnamefont{and}
  \bibinfo{author}{\bibfnamefont{M.}~\bibnamefont{Dettenmaier}},
  \bibinfo{journal}{Polymer} \textbf{\bibinfo{volume}{23}},
  \bibinfo{pages}{1305} (\bibinfo{year}{1981}).

\bibitem[{\citenamefont{Kausch et~al.}(1987)\citenamefont{Kausch, Petrovska,
  Landel, and Monnerie}}]{kausch87}
\bibinfo{author}{\bibfnamefont{H.~H.} \bibnamefont{Kausch}},
  \bibinfo{author}{\bibfnamefont{D.}~\bibnamefont{Petrovska}},
  \bibinfo{author}{\bibfnamefont{R.~F.} \bibnamefont{Landel}},
  \bibnamefont{and} \bibinfo{author}{\bibfnamefont{L.}~\bibnamefont{Monnerie}},
  \bibinfo{journal}{Polymer Engineering \& Science}
  \textbf{\bibinfo{volume}{27}}, \bibinfo{pages}{149} (\bibinfo{year}{1987}).

\bibitem[{\citenamefont{Benkoski et~al.}(2001)\citenamefont{Benkoski,
  Fredrickson, and Kramer}}]{benkoski01}
\bibinfo{author}{\bibfnamefont{J.~J.} \bibnamefont{Benkoski}},
  \bibinfo{author}{\bibfnamefont{G.~H.} \bibnamefont{Fredrickson}},
  \bibnamefont{and} \bibinfo{author}{\bibfnamefont{E.~J.}
  \bibnamefont{Kramer}}, \bibinfo{journal}{J Polym Sci Part B: Polym. Phys.}
  \textbf{\bibinfo{volume}{39}}, \bibinfo{pages}{2363} (\bibinfo{year}{2001}).

\end{thebibliography}
\bibliographystyle{apsrev}
\end{document}